\documentclass[a4paper,fleqn,usenatbib]{mnras}

\usepackage[T1]{fontenc}
\usepackage{ae,aecompl}

\usepackage{graphicx}	
\usepackage{amsmath}	
\usepackage{amssymb}	
\usepackage{color}
\usepackage{enumerate}
\usepackage{xspace}

\newcommand{\Nant}{\ensuremath{N_{\mathrm{a}}}\xspace}
\newcommand{\Ng}{\ensuremath{N_{\mathrm{g}}}\xspace}
\newcommand{\s}{\ensuremath{\hat{\mathbf{s}}}} 
\newcommand{\spix}{\ensuremath{\hat{\mathbf{s}}_{0}}}
\newcommand{\Cna}[1][n]{\ensuremath{\mathcal{C}^{(#1)}_{a,\spix}}}
\newcommand{\Kna}[1][n]{\ensuremath{\mathcal{K}^{(#1)}_{a,\spix}}}

\newcommand{\ra}{\ensuremath{\mathbf{r}_a}}
\newcommand{\rb}{\ensuremath{\mathbf{r}_b}}

\newcommand{\caliter}{400}
\newcommand{\damp}{\ensuremath{\gamma}}

\newcommand{\tcal}{\ensuremath{t_{\mathrm{cal}}}}
\newcommand{\teff}{\ensuremath{t_{\mathrm{eff}}}}

\title[EPICal]{An Efficient Feedback Calibration Algorithm for Direct Imaging Radio Telescopes
}

\author[Beardsley et al.]{
Adam P. Beardsley,$^{1}$\thanks{E-mail: Adam.Beardsley@asu.edu}
Nithyanandan Thyagarajan,$^{1}$
Judd D. Bowman$^{1}$
\newauthor
and Miguel F. Morales$^{2}$
\\
$^{1}$Arizona State University, School of Earth and Space Exploration, Tempe, AZ 85287, USA\\
$^{2}$University of Washington, Department of Physics, Seattle, WA 98195, USA\\
}

\date{Accepted XXX. Received YYY; in original form ZZZ}

\pubyear{2016}

\begin{document}
\label{firstpage}
\pagerange{\pageref{firstpage}--\pageref{lastpage}}
\maketitle

\begin{abstract}
We present the E-field Parallel Imaging Calibration (EPICal) algorithm, which addresses the 
need for a fast calibration method for direct imaging radio astronomy correlators. Direct 
imaging involves a spatial fast Fourier transform of antenna signals, alleviating 
an 
$\mathcal{O}(\Nant^2)$ computational bottleneck typical in radio correlators, and yielding a more gentle $\mathcal{O}(\Ng \log_2 
\Ng)$ scaling, where \Nant is the number of antennas in the array and \Ng is the number of grid points in the imaging analysis. This can save orders of magnitude in computation cost for next generation arrays 
consisting of hundreds or thousands of antennas. However, because antenna signals are mixed in the imaging
correlator without creating visibilities, gain correction must be applied prior to imaging, rather than on visibilities post-correlation. We develop the EPICal algorithm 
to form gain solutions quickly and without ever forming visibilities. This method scales as the 
number of antennas, and produces results comparable to those from visibilities. 
We use simulations to demonstrate the EPICal technique and study the noise properties of
our gain solutions, showing they are similar to visibility based solutions in realistic 
situations.
By applying EPICal to two seconds of Long Wavelength Array data we achieve a 65\% dynamic range
improvement compared to uncalibrated images, showing this algorithm is a promising
solution for next generation instruments.
\end{abstract}



\section{Introduction}
In order to satisfy the survey speeds required for precision cosmology as well as searches for 
fast radio transients, radio astronomy is undergoing a paradigm shift toward interferometers 
consisting of hundreds to thousands of small, widefield antennas. Many arrays with this design 
are already built or under construction including the Hydrogen Epoch of Reionization 
Array\footnote{http://reionization.org} (HERA), the Murchison Widefield Array (MWA; 
\citealt{tin13,bow13}), the Donald C. Backer Precision Array for Probing the Epoch of 
Reionization (PAPER; \citealt{par10}), the LOw Frequency ARray (LOFAR; \citealt{van13}), the 
Canadian Hydrogen Intensity Mapping Experiment (CHIME; \citealt{ban14}), the Long 
Wavelength Array (LWA; \citealt{ell13}), and the low frequency Square Kilometer Array (SKA1-
Low; \citealt{mel13}).

The most common radio correlator designs of today, the FX and lag correlators \citep{rom99},
cross-multiply the signals from all pairs of antennas.
This computation scales as the number of antennas squared, $\mathcal{O}(\Nant^2)$ 
\citep{bun04}. As the number of elements in future arrays grows, the computational cost will 
become prohibitively expensive, and exploring efficient correlator schemes is essential to 
enable next generation instruments \citep{lon00}. Meanwhile, radio transient monitoring 
requires access to high time and frequency resolution data to identify and characterize events 
such as fast radio bursts (FRBs, \citealt{lor07}), or to follow up gravitational wave candidates 
with radio observations \citep{abb16a,abb16b}. FRBs are relatively unexplored at low frequencies 
(< 1 GHz) \citep{row16, tin15, tro13}, but are expected to occur on timescales $\Delta t \sim$ 1--10~ms \mbox{\citep{tho13}}. 
Recording the full visibility matrix for $\Nant \gtrsim 10^3$ arrays at this timescale leads to 
extremely high data rates. 

Direct imaging correlators are a new variety of radio correlator which aim to alleviate both the 
computational strain of forming $\Nant^2$ correlations and the high data throughput associated 
with short timescale science. This is done by performing a spatial fast Fourier transform (FFT) 
to image the antenna signals, then squaring and averaging in time. This process scales as 
$\mathcal{O}(\Ng \log_2 \Ng)$, where \Ng~is the number of grid points in the FFT \citep{mor11,
 teg09, teg10}.
 The choice of grid points depends on the specific algorithm and implementation, and ultimately determines the extent and pixel size of the output image.
 For certain classes of telescopes, significantly those envisioned for next 
 generation cosmology experiments, the computational scaling is a large improvement over the $\Nant^2$ 
 scaling of traditional methods. Furthermore, for the same temporal resolution, the  
 output bandwidth will be lowered in cases where $\Ng < \Nant^2$.

A handful of prototype direct imaging correlators have been tested on arrays including the 
Basic Element for SKA Training II (BEST-2) array \citep{fos14}, the Omniscope \citep{zhe14}, 
and an earlier pulsar timing experiment at GHz frequencies \citep{oto94, dai00}. Each of these 
are examples of so-called FFT correlators -- a subclass of direct imaging correlators which rely 
on identical antennas with restricted placement, which allows the FFT to be performed without 
gridding. We recently released the E-field Parallel Imaging Correlator \citep[EPIC;][]{thy17}, 
which is a software implementation of the Modular Optimal Frequency Fourier \citep[MOFF;][]
{mor11} imaging algorithm. This architecture leverages the software holography/A-transpose 
framework to grid electric field data streams before performing the spatial FFT, allowing for an 
optimal map without placing constraints on array layout or requiring identical antennas 
\citep{mor09,bha08,teg97a}.

A challenge common to all direct imaging algorithms is calibration of the antenna gains. With a 
traditional cross correlator, visibilities are 
averaged in time, reducing the data volume, then used to calibrate 
after observing and before further processing such as imaging. However, a direct imaging correlator mixes 
the signals from all antennas before averaging in time, making calibration a 
requirement at the front end, before any averaging and imaging. Previous solutions have involved 
applying calibration solutions generated from a parallel FX correlator \citep{zhe14, fos14}, or 
integrating a dedicated FX correlator which periodically formed the full visibility matrix to solve 
for gains \citep{wij09,dev09}. While these solutions were sufficient to enable the exploration of 
FFT correlators and beamformers, they will not scale to future arrays with $\Nant \gtrsim 10^3$.

Here we present the E-field Parallel Imaging Calibration (EPICal) algorithm -- a novel solution 
to the calibration problem, which can be integrated into direct imaging correlators and scales 
only as the number of antennas, $\mathcal{O}(\Nant)$ (Section~\ref{sec:cal_derive}). This method uses a correlation of the 
uncalibrated antenna signal stream with an output image pixel from the backend of the 
correlator to solve for the complex gains of the antennas. Because the calibration must be 
applied before gridding and imaging, our solution requires an iterative approach where the data 
from one time series is used to update the gains which are applied to the following time series. 
An example implementation of the algorithm is available with the EPIC software 
package\footnote{http://github.com/nithyanandan/EPIC}.

This work is a direct extension of \citealt{thy17}. We
recommend the interested reader refer back to that work for details about the MOFF
algorithm itself and the EPIC implementation.
In this manuscript we review the MOFF algorithm and derive the calibration algorithm in \S \ref{sec:math}. We 
then demonstrate the algorithm in simulations in \S \ref{sec:sim}, 
and discuss noise trends in \S \ref{sec:noise}. We apply the algorithm to a sample LWA 
data set in \S \ref{sec:data}.
Finally we conclude and discuss potential extensions to the 
algorithm in \S \ref{sec:discussion}.

\section{Mathematical Framework}\label{sec:math}
\subsection{Review of the MOFF algorithm}
\begin{figure}
\begin{center}
\includegraphics[width=\columnwidth]{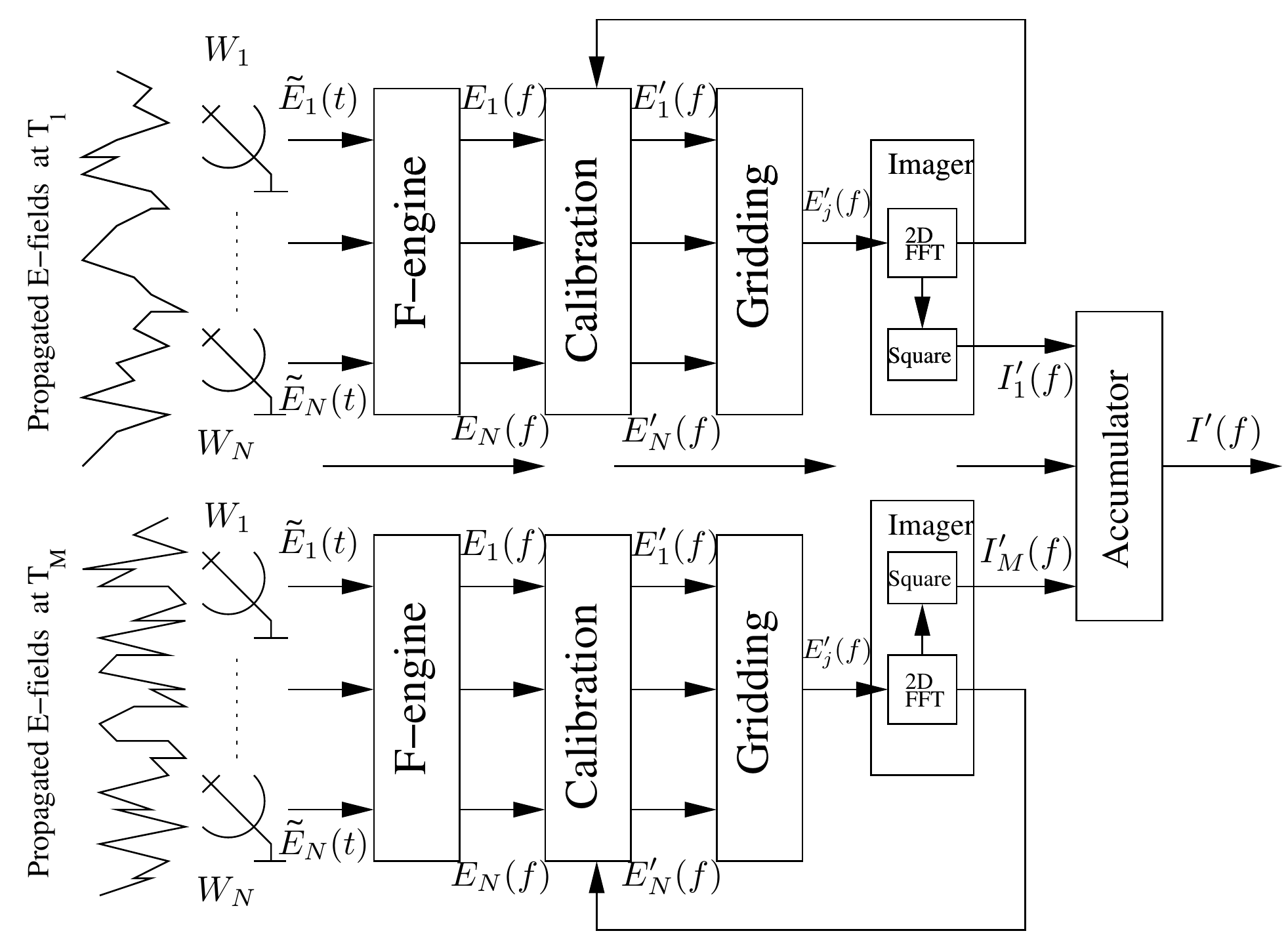}
\caption{Data flow for the MOFF algorithm, reproduced from \citealt{thy17} (their Fig. 1). The time domain
electric field measurements are collected by the antenna elements then frequency transformed in the 
F-engine. The calibration is applied, then the electric field measurements are gridded using the 
antenna aperture illumination patterns. The imager performs a spatial 2D FFT on the gridded data before 
squaring and averaging. The accumulated images are then written to disk.}
\label{fig:moff_flow}
\end{center}
\end{figure}

We begin by reviewing the data flow of the MOFF algorithm, highlighting aspects relevant to 
the calibration. The interested reader is encouraged to refer to \citealt{mor11} and \citealt{thy17} for a more 
thorough discussion. Figure~\ref{fig:moff_flow} is reproduced from \citealt{thy17} (their Fig. 1) to illustrate 
the various steps of the algorithm.

We begin by considering the electric field vector incident on the ground as a function of position and time, 
$\mathbf{\widetilde{E}}(\mathbf{r},\tau)$. Following \citealt{cla99}, we make a few assumptions in order to clarify notation and 
simplify the problem. 
First we note that the electric field over a finite time interval, $\Delta t$, can be expressed as a Fourier series with coefficients corresponding to frequencies $f$. 
The Fourier coefficients are related to the time-domain electric field by\footnote{
A digital system will discretize the time-domain electric field and estimate 
equation~\ref{eq:spectrum} as a sum for a finite number of frequency channels. At this stage 
we are considering the true electric field and keep the integral.
}\citep[e.g.][Appendix 3.1]{tho01}
\begin{equation}\label{eq:spectrum}
\mathbf{E}(\mathbf{r},f_j,t) = \frac{1}{\Delta t} \int_{t-\Delta t/2}^{t+\Delta t/2} \mathbf{\widetilde{E}}(\mathbf{r},\tau) e^{-2 \pi i f_j \tau} d\tau.
\end{equation}
Because we are considering a finite time interval, the coefficients are independent
for discrete frequencies, $f_j = j/\Delta t$, for all integers $j$. 
Likewise, the variable $t$ represents discrete times separated by $\Delta t$.
The time-domain electric field over the same time interval, $t-\Delta t/2 \le \tau \le t+\Delta t/2$,
can be recovered by an infinite sum,
\begin{equation}
\mathbf{\widetilde{E}}(\mathbf{r},\tau)  = \sum_{j=-\infty}^{\infty} \mathbf{E}(\mathbf{r},f_j,t) e^{2 \pi i f_j \tau} .
\end{equation}
By analyzing the quasi-monochromatic components of the electric field spectrum,
$\mathbf{E}(\mathbf{r},f_j,t)$, we can treat frequency channels independently. For brevity, 
henceforth we will omit the subscript $j$ from the frequency variable.

We will assume all antennas lie in a plane, 
reducing the problem to two dimensions. 
The MOFF algorithm in principle can account for non-coplanar arrays through
the W-projection algorithm \citep{cor08} by using the electric field Fresnel diffraction pattern
to project each antenna measurement to a common plane.
Because the EPIC code does not 
currently support this, we leave it to future work.

Next we treat the electric field for each polarization separately. This allows us 
to express the electric field for each polarization as a scalar, $E(\mathbf{r},f,t)$, and treat subsequent field 
propagation and measurements as scalar rather than tensor multiplications. A full polarization 
treatment is required to construct polarized sky images, however the calibration will only 
require a sky model expressed in the instrumental polarization basis. We can therefore treat 
each instrument polarization independently.

Finally we assume the sources of the electric field are far away, allowing us to describe the 
electric fields propagating from sky sources as originating from an imaginary celestial sphere 
located a large distance from the ground. We express this intermediary electric field as an 
angular distribution, $\mathcal{E}(\hat{\mathbf{s}},f,t)$, where $\hat{\mathbf{s}}$ is the unit 
vector in the direction of a patch of the celestial sphere. We further assume the space between 
the celestial sphere and the ground is empty, allowing us to express the electric field incident 
on the ground as a Fourier integral of the propagated fields across the hemisphere above the horizon.
\begin{equation}\label{eq:sky_propagated}
E(\mathbf{r},f,t) = \int_{\mathrm{sky}} \mathcal{E}(\hat{\mathbf{s}},f,t) e^{-2\pi i f\mathbf{r} \cdot \hat{\mathbf{s}}/c}\, d\Omega
\end{equation}  
Here $d\Omega$ is the differential solid angle.
By making the latter assumptions we have left treatment of the ionosphere and terrestrial 
emission such as local interference or thermal radiation from the ground for future work. 

Next we turn to the measurement performed by an antenna with frequency dependent far-field radiation pattern 
$\mathcal{W}_a(\hat{\mathbf{s}},f)$, where the subscript $a$ denotes a specific antenna. 
The far-field radiation pattern of the antenna gives the direction-dependent response 
of the antenna, and the total signal can be taken as the sum over all directions 
\citep[e.g.][]{kra86}.
This allows us to express the electric field measurement that can be recovered from the 
induced voltage on an ideal antenna as a weighted
integral over the sky \citep[e.g.][]{teg10}.
\begin{equation}
E^I_a(f,t) = \int_{\mathrm{sky}}  \mathcal{W}_a(\hat{\mathbf{s}},f)\mathcal{E}(\hat{\mathbf{s}},f,t) e^{-2\pi i f\mathbf{r}_a \cdot \hat{\mathbf{s}}/c}\, d\Omega. 
\end{equation}
Here we have adopted the common convention of normalizing the 
radiation pattern to have response of unity at the center of the main lobe \citep[e.g.][]{nap99}.

Equivalently, we can employ properties of the Fourier transform to express the effect of the antenna radiation pattern in the plane of the 
aperture.
\begin{equation}
E^I_a(f,t) = \int_{\mathrm{aperture}}W_a(\mathbf{r}-\mathbf{r}_a,f)E(\mathbf{r},f,t)\,d^2\mathbf{r}
\end{equation}
Here we used the Fourier relationship between the antenna's far-field radiation pattern and the 
aperture illumination pattern (sometimes referred to as the aperture distribution), 
\begin{equation}
W_a(\mathbf{r}-\mathbf{r}_a,f)= \frac{f^2}{c^2}\int_{\mathrm{sky}}   \mathcal{W}_a(\hat{\mathbf{s}},f) e^{2\pi i f(\mathbf{r}-\mathbf{r}_a) \cdot \hat{\mathbf{s}}/c}\, d
\Omega.
\end{equation} 
Intuitively this is the response function by which the antenna weights the electric
fields over the aperture, and sums at the feed.
While it is possible to define each antenna aperture illumination pattern in a way to absorb the 
relative position of the antenna, we keep the $\mathbf{r}_a$ coordinate to retain the separable
concepts of the primary beam (the product of two antenna radiation patterns) and the array
point spread function (square of the Fourier transform of the array layout).

Thus far we have considered the perfectly measured electric field distribution according to 
the antenna far-field radiation patterns. Here we introduce a multiplicative complex gain as 
well as an additive noise term which corrupt the measured signals at each of the antennas.
\begin{equation}\label{eq:apply_gain}
E_a(f,t) = g_a(f,t) E_a^I(f,t) + n_a(f,t)
\end{equation}
We have carried the frequency dependence to this point to make clear the complex gains are
frequency dependent. However, each subsequent step treats each frequency 
channel independently, so we will drop the $f$ to simplify notation. 
Furthermore, while antenna gains may vary with time, we will assume they are constant over
the period of our calibration, and thus drop their time dependence from the notation.

The first step of the MOFF algorithm is to digitize the antenna signals and channelize the data over a finite time interval, effectively estimating the electric field spectrum 
in equation~\ref{eq:spectrum} over a finite frequency bandwidth.
This operation is often referred to as 
the F-engine of a correlator. The MOFF shares the same F-engine design as an FX cross
correlator which performs the frequency transform (F-engine) before cross multiplying 
(X-engine). For a recent review of spectrometers used in radio astronomy, please see \citealt{pri16b}.

The MOFF algorithm next calls for a calibration of the electric field measurements. The goal of our new calibration method will be 
to form an estimate of the gains, $g'_a$, where the prime represents an estimate. For now we will assume we have formed an estimate 
to proceed with the MOFF pipeline. We correct the incoming electric field data stream using our 
current estimate of the gains.
\begin{equation}
E'_a(t) = E_a(t)/ g'_a
\end{equation}

The next step is to grid the calibrated electric field measurements to a finite set of regularly spaced 
positions on the plane of the ground, $\mathbf{r}_j$.
The extent of the grid is chosen to encompass the footprint of the antenna array.
The spacing of grid points should be chosen to be smaller than the antenna size to achieve sufficient sampling, and a typical choice is several grid points per linear dimension of the antennas.
The size of the grid spacing determines the field of view of the image (through an
inverse relationship), so this choice is typically limited on one end by the antenna size 
(instrumental field of view), and on the other by half the observation wavelength (full sky).
The EPIC implementation currently only supports a rectangular grid,
but in principle the type of grid can be extended to any arrangement which can be Fourier
transformed with $\mathcal{O}(\Ng \log_2 \Ng)$ computational complexity including
hexagonal grids \citep{mer79} or arbitrary hierarchies of rectangular grids with shears and rotations at each level \citep{teg10}.

MOFF uses the antenna aperture illumination patterns as the gridding 
kernel according to the software holography/A-transpose technique \citep{mor09,bha08}. 
Mathematically we can express the gridding operation as
\begin{equation}\label{eq:gridding}
E'_j=\sum_a W_a(\mathbf{r}_j-\mathbf{r}_a) E'_a(t)
\end{equation}
Formally the value at each pixel is calculated by summing the antenna
measurements over all antennas, each weighted by its respective aperture illumination
pattern.
However the aperture illumination patterns are typically compact and non-overlapping, so most pixels
will only contain contribution from one antenna, allowing equation~\ref{eq:gridding}
to be calculated through a sparse matrix multiplication.
This gridding 
places the data on a regular grid and achieves an
optimally weighted map, analogous to widely used Cosmic Microwave Background analysis techniques \citep{teg97b}. Unlike 
traditional correlators which grid visibilities, the MOFF grids the electric field measurements directly. This 
operation must be performed for every time interval, before any averaging of the data.

After gridding, the imaging portion of the MOFF performs a 2D spatial FFT to form 
instantaneous electric field images of the sky at fixed locations $\s_k$.
The FFT is possible because we have gridded to regularly spaced samples in a 2D plane.
The Fourier dual coordinates to $\mathbf{r}f/c$ are $(l,m)$, where $l=\sin\theta\cos\phi$,
$m=\nobreak\sin\theta\sin\phi$, and $(\theta,\phi)$ are the zenith 
and azimuthal angles\footnote{Observations pointed away from zenith can be imaged by applying a phase to the electric fields in the same way one would phase visibilities, and redefining $(\theta, \phi)$ coordinates relative to the phase center.}. 
The full unit vector towards the sky is 
$\s=(l,m,\sqrt{1-l^2-m^2})$.
Being direction cosines, grid locations with ${l^2+m^2>1}$ are non-physical and should
be ignored.

The algorithm up to this point can be summarized 
through the following equation.
\begin{align}
\mathcal{E}'(\s_k,t) & = \underbrace{\frac{1}{\Nant} \sum_j e^{2\pi i f\mathbf{r}_j \cdot \s_k/c}}_{\mathrm{2D\;FFT}} 
\underbrace{\sum_a W_a(\mathbf{r}_j - \ra)}_{\mathrm{Gridding}} \\
& \qquad \times \underbrace{ \frac{1}{g'_a} \left(g_a E_a^I(t)+n_a(t)\right)}_{\mathrm{Calibration}} \nonumber
\end{align}
The sum over antennas uses the antenna aperture illumination pattern to grid the calibrated electric field measurements onto 
regular gridpoints, $\mathbf{r}_j$. The sum over $j$ denotes the 2D FFT to sky coordinates, resulting in an 
estimate for the instantaneous electric field image. 
The sums over $a$ and $j$ combine the antenna signals, which is the primary 
reason calibration must be performed beforehand where individual antennas retain their 
identities.

As an illustrative step, we can simplify the above expression for two regimes:
where the grid size is significantly smaller than the antenna size, or all antennas are identical
and pseudo-randomly placed (or both). In these cases, after exchanging the sums,
we can use
the discrete Fourier transform to approximate the Fourier relationship between the antenna
aperture illumination pattern and the far-field radiation pattern.
\begin{align}\label{eq:epix}
\mathcal{E}'(\s_k,t) & = \frac{1}{\Nant} \sum_a \frac{1}{g'_a}\left(g_a E^I_a(t)+n_a(t)\right) e^{2\pi i f\s_k \cdot \ra/c} \nonumber \\
  & \qquad\times \sum_j W_a(\mathbf{r}_j-\ra)e^{2\pi i f\s_k \cdot (\mathbf{r}_j-\ra)/c} \nonumber\\
& \approx \frac{1}{\Nant} \sum_a \frac{1}{g'_a}\left(g_aE^I_a(t)+n_a(t)\right) e^{2\pi i f\s_k \cdot \ra/c}\mathcal{W}_a(\s_k)
\end{align}
While this step is not necessary (and in fact is not a part of the imaging algorithm), 
the approximation holds for most current and
planned cosmology experiments, so we will proceed with the more simple expression.
In this form we see that the effect of gridding with the antenna aperture illumination pattern is to attenuate 
the image by a factor of the far-field radiation response, $\mathcal{W}_a(\s)$.
This added factor of the instrument response is sometimes referred to as the
\emph{holographic frame} in the map-making literature \citep[e.g.][]{mor09, sul12}.

Finally, the images are squared and averaged in time to form dirty images, 
$I'(\hat{\mathbf{s}}_k)$, then written 
to disk. While operations have been reordered, the resulting image is equivalent to one 
produced by gridding and imaging visibilities, up to a precision set by gridding coarseness. By shifting from cross-correlation of all pairs of 
antennas to a spatial FFT, the computational cost is significantly lowered for certain classes of 
arrays, particularly those with densely-filled apertures of small antennas. For these arrays,
the output bandwidth is also greatly reduced as visibilities are replaced with
gridded images. These gridding, computational, and data bandwidth differences were explored in \citealt{thy17} (sections 4.3, 6.1, and 6.2 respectively).

\subsection{Derivation of calibration}\label{sec:cal_derive}
We saw above that the MOFF algorithm (in general any direct imaging architecture) combines the 
signals from all antennas, and thus requires the calibration to be applied 
before imaging. Furthermore, visibilities are never formed, which are 
traditionally the basic measurement used to form calibration solutions. Here we derive an 
alternative method to estimate the antenna gains quickly, using the data products of the 
MOFF. 

\begin{figure}
\begin{center}
\includegraphics[width=\columnwidth]{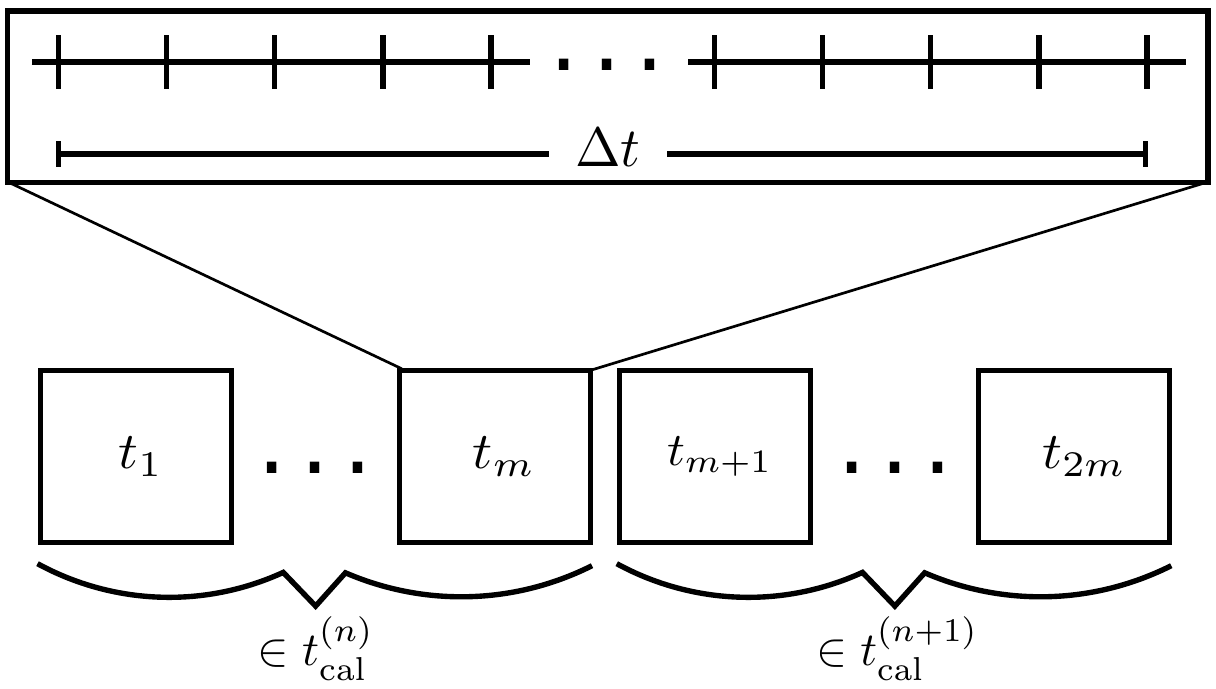}
\caption{Schematic of timescales involved in the calibration algorithm. The top box depicts
several digitized time samples of the observation. 
It is common that real time samples are separated by the Nyquist interval of $1/2B$ (or $1/B$ for complex samplers), where $B$ is the bandwidth of the observation.
The antenna electric field spectra, $E_a(f,t)$, are computed over the time interval $\Delta t$ for consecutive times as shown. 
A finite stream of these intervals, labeled by the set of times $t^{(n)}_{\mathrm{cal}}$, will be 
used to generate an $n^{\mathrm{th}}$ estimate of the gains, which are applied to the 
following stream, labeled by $t^{(n+1)}_{\mathrm{cal}}$. The gains are assumed to be constant between the two time streams $t^{(n)}_{\mathrm{cal}}$ and $t^{(n+1)}_{\mathrm{cal}}$.
}
\label{fig:times}
\end{center}
\end{figure}

For some applications, the electric field measurements can be recorded
directly, and gains may be determined in post-processing, such as in our demonstration in 
Section~\ref{sec:data}. However, with the large-\Nant,
high bandwidth, and high duty-cycle instruments required for future cosmological 
experiments, it is only practical to record time averaged quantities. Therefore we will assume
a requirement that gains must be determined on relatively short timescales equal to or less than the 
stability timescale of the instrument.
We will adopt a strategy to use a finite stream of data to form a gain estimate, which we will
apply to the subsequent stream of data. 
This inherently assumes that the changes in gains are negligible over adjacent
time streams.
Figure~\ref{fig:times} illustrates the several timescales involved in the calibration algorithm.

In the case of exceptionally stable instruments, where gains are constant over long 
timescales, the requirement for a fast calibration scheme is alleviated and more sophisticated
fitting strategies can be undertaken, as we briefly discuss in Section~\ref{sec:discussion}.
Even in this ideal case, we will avoid adding to the computational strain of the
correlator by using quantities easily accessible to a direct imaging system.

Because data from one interval is used to update the gain solutions for the next, the 
process of finding a solution is an iterative one. We will use parenthetical superscripts to 
denote the calibration loop number. For example, we will assume we have already formed an 
estimate from $n$ loops, $g^{(n)}_a$. We will use these estimates to calibrate the next stream 
of data in order to form an updated estimate, $g^{(n+1)}_a$.

As a starting point, we consider the feedback calibration outlined in \citealt{mor11}. There it 
was suggested to form a correlation of the uncalibrated antenna measurements with the complex conjugate of an image 
pixel from the output of the correlator. 
We will see below that this quantity can be 
related to the sum of visibilities involving the antenna used in the correlation -- exactly the sum 
needed to calibrate in the case of a single point source sky. However, we aim for a more 
generalized solution for arbitrarily complex sky models. We therefore study the full expression 
for the antenna-pixel correlation,
\begin{equation}\label{eq:Kna_def}
\Kna \equiv \left<E_a(t) \mathcal{E}'^*(\spix,t)\right>_{t^{(n)}_{\mathrm{cal}}},
\end{equation}
where the superscript $n$ again represents the quantity formed in the $n^\mathrm{th}$ 
calibration loop,
and we use $t^{(n)}_{\mathrm{cal}}$ outside the brackets on the right-hand side to indicate 
that averaging is over all $t$ within the $n^\mathrm{th}$ calibration loop (see figure~\ref{fig:times}).
The superscript $^*$ indicates complex conjugation,
and \spix\, is the pixel center nearest a bright calibrator of interest. The 
following will hold for any chosen pixel, \spix, though it is advantageous to choose a pixel which 
contains a bright source to achieve a high signal to noise ratio. 

Plugging equation \ref{eq:epix} into equation \ref{eq:Kna_def}, 
and substituting the $g^{(n)}_b$ for $g'_b$ to indicate the specific loop number,
we find,
\begin{align}
\Kna & = \left<\left(g_a E_a^I(t)+n_a(t)\right) \frac{1}{\Nant} \right. \nonumber \\
& \qquad \left. \times \sum_b \frac{1}{g^{*(n)}_b}\left(g^*_bE_b^{I*}(t)+n^*_b(t)\right) \right.\nonumber \\
& \qquad \left. \times e^{-2\pi i f\spix \cdot \rb/c} \mathcal{W}^*_b(\spix)\right>_{t^{(n)}_{\mathrm{cal}}} \nonumber \\
& = \frac{1}{\Nant} \sum_b \frac{1}{g^{*(n)}_b} \mathcal{W}^*_b(\spix) e^{-2\pi i f\spix \cdot \rb/c} \nonumber \\
& \qquad \times \left<\left(g_aE_a^I(t)+n_a(t)\right)\left(g^*_bE_b^{I*}+n^*_b(t)\right) \right>_{t^{(n)}_{\mathrm{cal}}} \nonumber \\
& = \frac{1}{\Nant} \sum_b \frac{1}{g^{*(n)}_b} \mathcal{W}^*_b(\spix) e^{-2\pi i f\spix \cdot \rb/c} \nonumber \\
& \qquad \times \left(g_a g^*_b \left<E^I_a(t)E^{*I}_b(t)\right>_{t^{(n)}_{\mathrm{cal}}} + g_a\left<E_a^I(t)n^*_b(t)\right>_{t^{(n)}_{\mathrm{cal}}} \right. \nonumber \\
& \qquad \left. + g^*_b\left<n_a(t)E^{I*}_b(t)\right>_{t^{(n)}_{\mathrm{cal}}} + \left<n_a(t) n^*_b(t)\right>_{t^{(n)}_{\mathrm{cal}}}\right)  
\end{align}
where in the second step we group time-dependent terms, 
and in the third step we expanded the product within the time average.
Because the noise does not correlate with the ideal antenna measurements we can
drop the middle terms. We will also assume that the cross-correlation noise ($a\ne b$) 
is zero mean. Then by defining an ideal visibility as
$V_{ab}^I \equiv \left<E^I_a(t)E^{*I}_b(t)\right>_t$, we can simplify to
\begin{align}\label{eq:Kna}
\Kna & = \frac{1}{\Nant} \sum_b \frac{1}{g^{*(n)}_b} \mathcal{W}^*_b(\spix) e^{-2\pi i f\spix \cdot \rb/c}\nonumber \\
& \qquad \times \left( g_ag_b^* V_{ab}^I + \delta_{ab}\left<|n_a(t)|^2\right>_{t^{(n)}_{\mathrm{cal}}}\right)
\end{align}
where $\delta_{ab}$ is the Kronecker-delta function. To be clear, visibilities are never directly 
formed from the data. Rather it is convenient to express $\Kna$ in terms of the ideal 
visibilities, which we can model when solving for the gains below.

We will ultimately wish to model the right side of equation~\ref{eq:Kna} in order to solve for the
antenna gains. However, the self-correlated noise, $\left<|n_a(t)|^2\right>$, can often be 
significantly larger than the cross correlation power, and difficult to model precisely. It 
is therefore beneficial to subtract the self-correlation directly from $\Kna$. By properly
accounting for the various weighting and phasing terms, we can use self-correlations of the
uncalibrated antenna electric field measurements to remove the corresponding
terms from equation~\ref{eq:Kna}:
\begin{align}\label{eq:Cna}
\Cna & \equiv \Kna - \frac{1}{\Nant g^{*(n)}_a}\mathcal{W}^*_a(\spix)e^{-2\pi i f\spix \cdot \ra/c} \left<|E_a|^2\right>_{t^{(n)}_{\mathrm{cal}}}\nonumber \\
& = \frac{g_a}{\Nant} \sum_{b\ne a} \frac{g^*_b}{g^{*(n)}_b} \mathcal{W}^*_b(\spix) e^{-2\pi i f\spix \cdot \rb/c} V^I_{ab}.
\end{align}
Due to the above subtraction, the sum is now over $b \ne a$, explicitly removing the
dependence on the self-correlated antenna signals.

The next step is to use this antenna-pixel correlation to update our gain solution. We will model 
the right hand side of equation~\ref{eq:Cna} by assuming our current gain estimates are 
approximately correct, $g^{(n)}_b\approx g_b$, and replacing the ideal visibilities with a set of 
model visibilities, $V^M_{ab}$. These model visibilities rely on knowledge of the sky, and should 
be precomputed before observing. We can then solve equation~\ref{eq:Cna} for $g_a$ to achieve an 
updated estimate.
\begin{equation}\label{eq:cal_solution}
g^{(n+1)}_a = \frac{\Cna \Nant }{ \sum_{b\ne a} \mathcal{W}^*_b(\spix) e^{-2\pi i f\spix \cdot \rb/c} V^M_{ab} }
\end{equation}
This equation is our prescription for estimating the antenna gains of a direct imaging array. The 
computation complexity of the part of the calculation that must be performed for 
each time interval
scales as $\mathcal{O}(\Nant)$ since we form 
an antenna-pixel correlation, $\Cna$, for each 
antenna. 

\begin{figure}
\begin{center}
\includegraphics[width=\columnwidth]{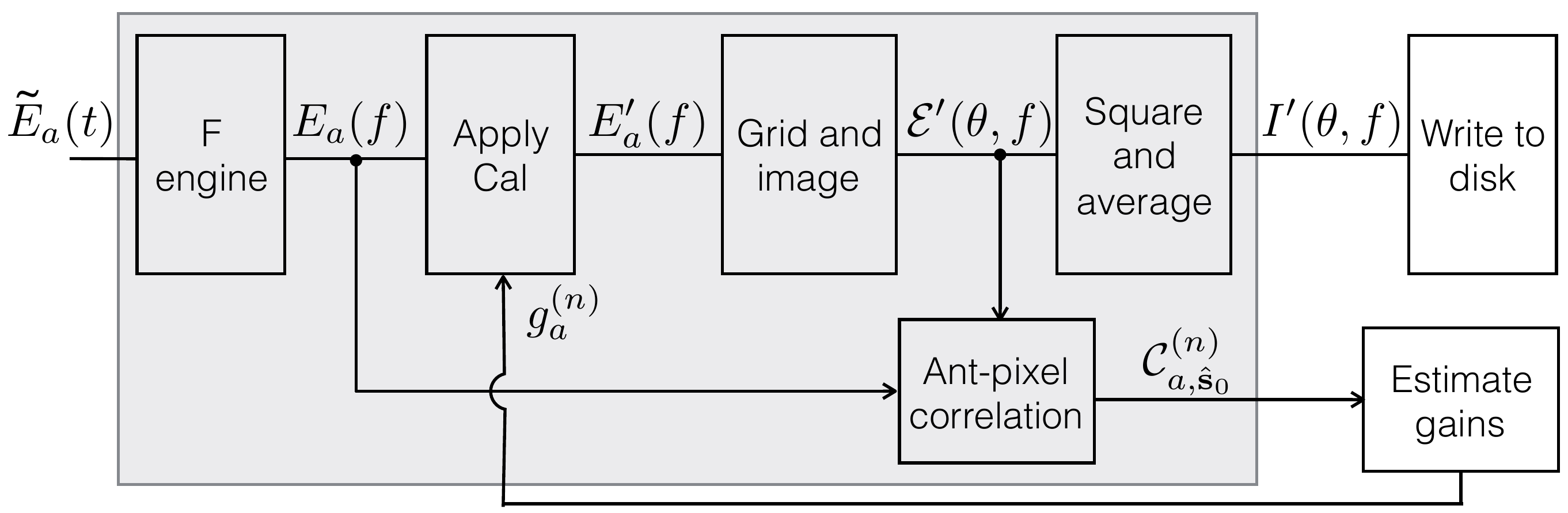}
\caption{The general data flow of the MOFF correlator, with a feedback calibration loop. A pixel 
from the (unsquared) image is tapped out and correlated against the input antenna electric field 
measurements to form $\Cna$ coefficients (equation~\ref{eq:Cna}). These coefficients are then 
used to update the gain estimates using equations~\ref{eq:cal_solution}~and~\ref{eq:damping},
and any additional fitting (Section \ref{sec:discussion}).
The gray box shows operations which must be done at high speed (before averaging in time),
while the white boxes show operations which can be performed at lower speeds because
they involve time averaged quantities, and need to be updated on slower timescales.
}
\label{fig:schematic}
\end{center}
\end{figure}

We show schematically the process of calibrating a direct imaging correlator in 
Fig.~\ref{fig:schematic}. Computationally expensive steps that must be performed for each time interval
are shown inside the gray box. The uncalibrated antenna signals are tapped out after the 
F-engine and correlated against the output image pixel of interest. The correlated values are then used to estimate the gains
using equation~\ref{eq:cal_solution}, and additional fitting if desired 
(see Section \ref{sec:discussion}).
This step does not have
the stringent cadence requirement of the steps in the gray box because the 
antenna-pixel correlations have been 
averaged in time.
The gains are then passed back 
to the correlator to update the calibration for subsequent integration intervals. 

While testing we found equation~\ref{eq:cal_solution} resulted in oscillatory gain solutions over time, 
as is often the case in iterative methods. To mitigate this we 
introduce a damping factor, $0 \leq \damp <1$, which is used to attenuate the gain update, 
effectively giving the solutions memory of previous iterations.
\begin{equation}\label{eq:damping}
g^{(n+1)}_a = \frac{(1-\damp)\Cna \Nant }{ \sum_{b\ne a} \mathcal{W}^*_b(\spix) e^{-2\pi i f\spix \cdot \rb/c} V^M_{ab} }+ \damp g^{(n)}_a
\end{equation}
We found that while equation~\ref{eq:cal_solution} does indeed converge on good solutions, 
the process is made faster by tuning the damping factor. 
 Once the loop converges the damped version is 
essentially a weighted average over the past several iterations, giving the solutions a longer effective 
integration time. 

We conclude this section by connecting our calibration expression to that found in a visibility 
framework. In the special case of a single bright calibrating source at zenith, we can greatly 
simplify equations~\ref{eq:Cna} and~\ref{eq:cal_solution}. We will assume the radiation patterns are 
normalized such that $\mathcal{W}(0)=1$. We can further drop the exponential phase terms 
because $\spix$ is perpendicular to our planar antenna array. We then absorb the true gains into the true visibilities in 
equation~\ref{eq:Cna} to express as a sum of measured, uncalibrated, visibilities.
\begin{equation}
\mathcal{C}^{(n)}_{a,0} \rightarrow \frac{1}{\Nant}\sum_{b\ne a} \frac{1}{g^{*(n)}_b} V_{ab}
\end{equation}

We next plug this expression into equation~\ref{eq:cal_solution} to find our simplified 
calibration solution for a single bright point source. Because our sky is a single bright point 
source, the model visibilities are simply the flux of the source, $S_{\mathrm{src}}$.
\begin{equation}
g^{(n+1)}_a \rightarrow \frac{\sum_{b\ne a}  V_{ab}/g^{*(n)}_b}{\Nant S_{\mathrm{src}}}
\end{equation}
This is simply a gain-weighted sum of the measured visibilities over the flux of the source, 
which is indeed the limiting result from a visibility approach, for example seen in \citealt{mit08}. 
The ability to recover the equivalent expression despite not actually forming the visibilities is a 
result of the fact that only sums over visibilities come into the visibility-based solution, as was described in 
\citealt{mor11}. We have confirmed the limiting case equivalence here, and will explore the 
more general case in more detail in the following sections.

\section{Simulation}\label{sec:sim}
We first demonstrate our calibration method through a controlled simulation. A complex gain is 
created for each antenna with random phase and amplitude, which is used to corrupt the 
simulated data stream, then we attempt to estimate the gains using our calibration routine. 
We use the simulation software included in the EPIC package to create stochastic antenna measurements of a sky model.
In this and following sections we restrict ourselves to arrays of identical antennas,
and leave a demonstration with heterogenous arrays to future work.

\begin{figure}
\begin{center}
\includegraphics[width=0.9\columnwidth]{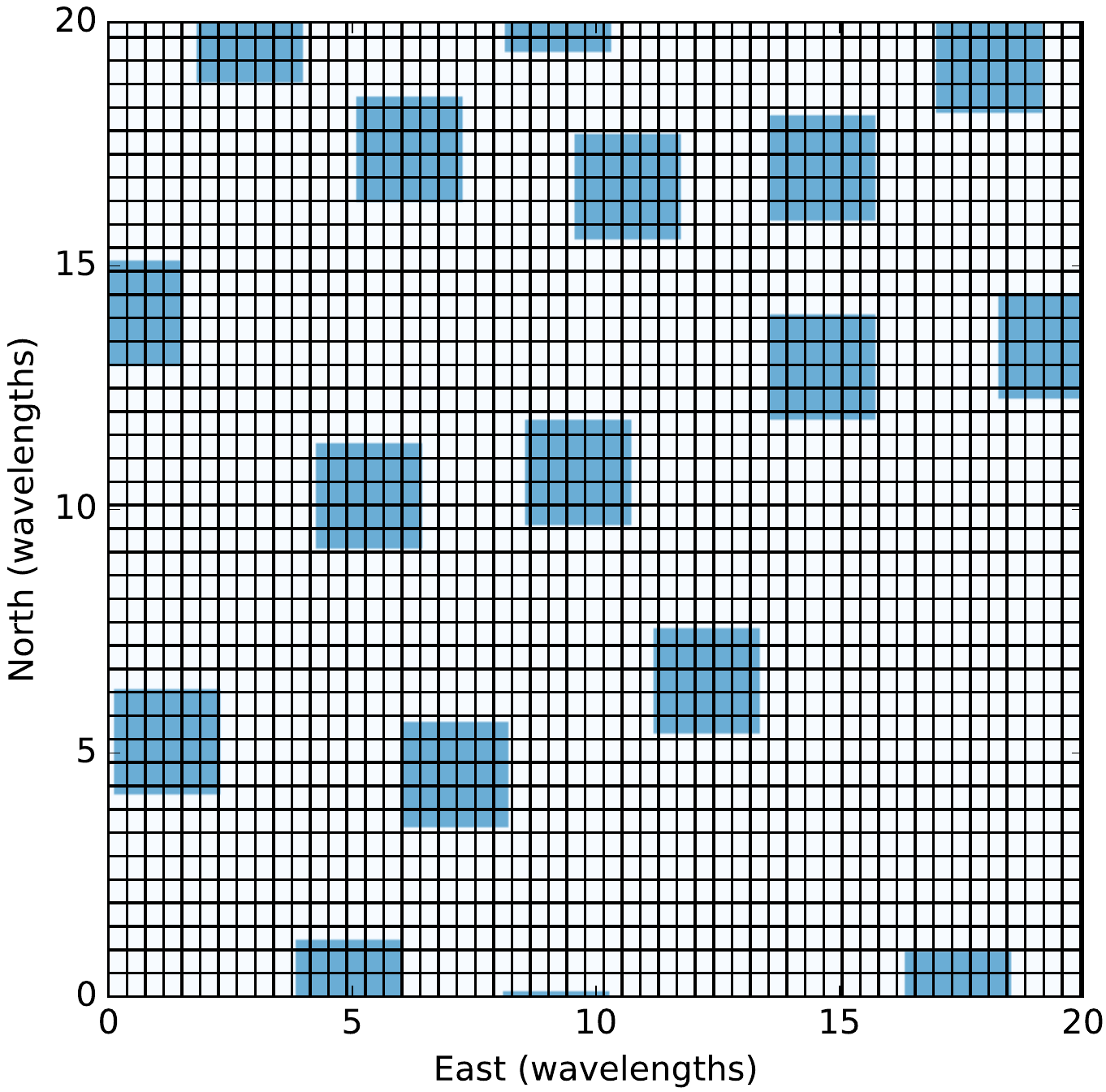}
\caption{A subsection of the simulated array in Section~\ref{sec:sim}, with overlaid 
grid. The blue squares show the locations of some of the antennas simulated. Grid points
used to grid the electric field measurements (equation \ref{eq:gridding}) are located at the intersections of black lines. Only a subsection of the array is shown so the grid points can be seen. We specified that the grid spacing be less than 0.5~wavelengths, and EPIC finds a grid which encompasses the array with the number of pixels in each dimension a power of two. This results in a grid spacing of 0.376 in the East-West direction, and 0.48 in the North-South direction, and a total grid size of $256\times256$ pixels.
}
\label{fig:grid}
\end{center}
\end{figure}

Our simulated antenna 
array consists of the inner 51 antennas of the MWA layout \citep{bea12}, within a bounding 
box of 150~m. 
The physical parameters of our simulation can be arbitrarily scaled to other cases. However, 
we provide typical units for an MWA observation as an example. 
The antenna aperture illumination pattern used is a 4.4~m square tophat; a very rough 
approximation to the square shaped tiles of the MWA. Because our algorithm treats frequency 
channels independently, we simulate only one channel. For context we treat this as a channel
at 150~MHz with 40 kHz bandwidth, and adopt a sampling period of 25~$\mu$s. 
A subsection of the array is shown in Figure~\ref{fig:grid}, along with lines depicting
the grid points used to grid the electric field measurements (equation~\ref{eq:gridding}).
The simulated signal 
consists of 10 random point sources with flux densitities 0.5~Jy~$\lesssim S \lesssim$~1~Jy 
within the main lobe of the primary beam.

For our unknown gains, we create a set of random complex numbers where the amplitude is 
drawn from a gaussian distribution centered around unity with width 0.25, and truncated at zero. The gain phases are drawn from a uniform distribution in the range $[-\pi, \pi)$. 
These are our ``true gains", and we apply them to the frequency
domain simulated antenna electric field measurements as in equation~\ref{eq:apply_gain}. Our analysis is 
blind to these values until the end of the process to check accuracy. The gain estimates are 
initialized with unity, $g^{(0)}_a=1$.

We next process and image \caliter~time samples (10~ms). We also form the antenna-pixel correlations, 
\Cna[0], used in our calibration loop. The pixel used for the correlation is the source with the 
largest apparent flux (intrinsic flux attenuated by the primary beam). These correlation values 
are used to update the gain estimates. We create perfect model 
visibilities of our 10 simulated sources to be used in equation~\ref{eq:damping}. 
The updated gain estimates are used to calibrate the following 
\caliter~time samples. Through trial and error we found a damping factor of $\damp=0.35$ 
resulted in the quickest convergence in this simulation.

\begin{figure}
\begin{center}
\includegraphics[width=\columnwidth]{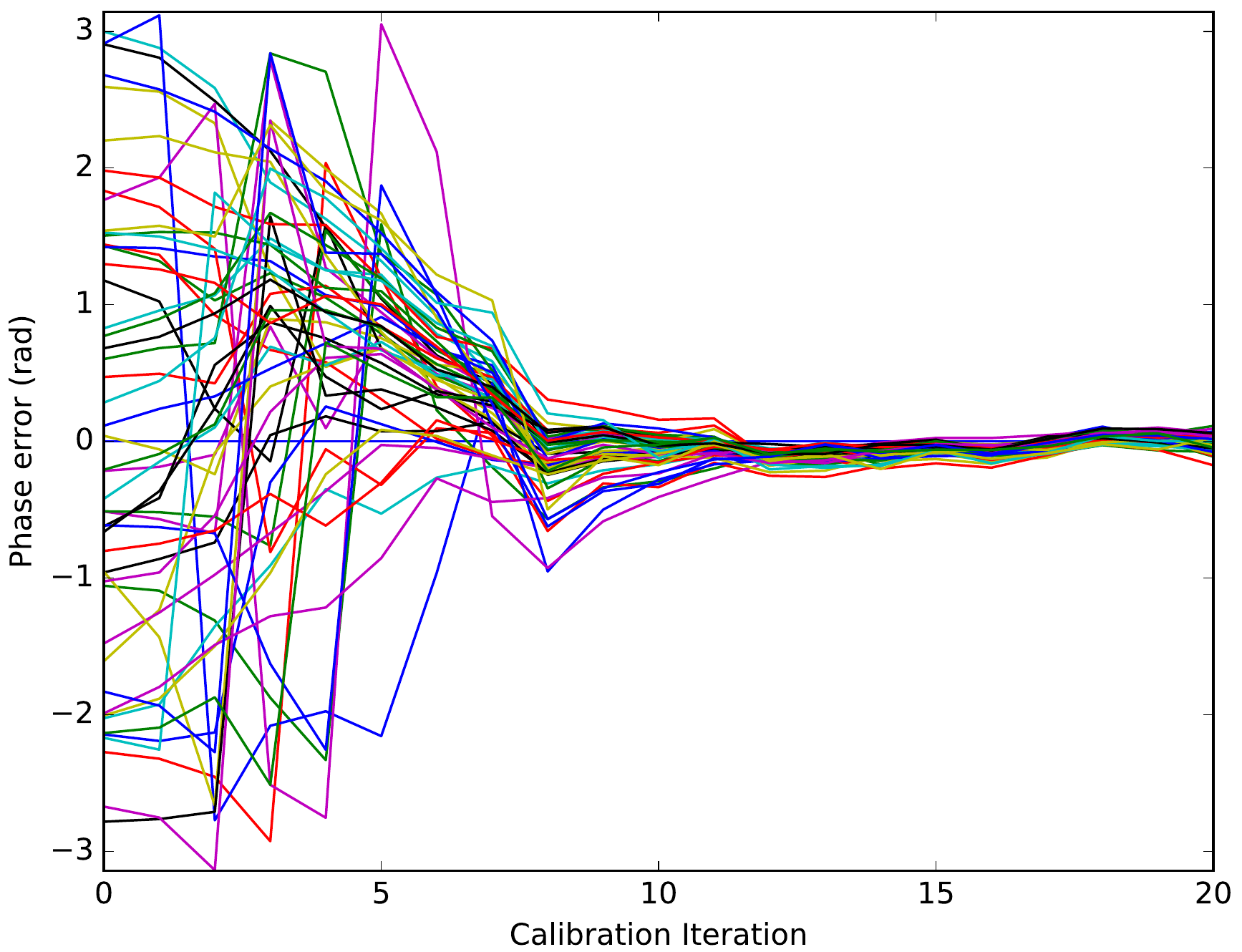}
\caption{Phase error of gain estimates as a function of iteration for simulated calibration. The 
gains were initialized with random phases, but the calibration loop was able to recover the 
correct phases after about 12 iterations. Each line represents an antenna in our 51 MWA 
antenna sample.
}
\label{fig:sim_phase}
\end{center}
\end{figure}

The calibration loop continues by updating the gain estimates every \caliter~time samples. The 
phase errors of our gain estimates are shown in Fig.~\ref{fig:sim_phase} for 20 such iterations. 
The phase error plotted is the phase relative to the true gain for each antenna (various colored 
lines). One antenna was used as a reference to fix the absolute phase, so has zero phase 
error. The other 50 antennas are shown to have error spanning the range $[-\pi,\pi)$ initially, and after about 
10 iterations lock into a solution, settling down to noise levels around iteration 12 (0.12~s). We 
stop the simulation when the updated gains trace the thermal noise of the simulated sources, 
which can be seen by the coherence of the 50 antenna gains after iteration 12.

\begin{figure}
\begin{center}
\includegraphics[width=\columnwidth]{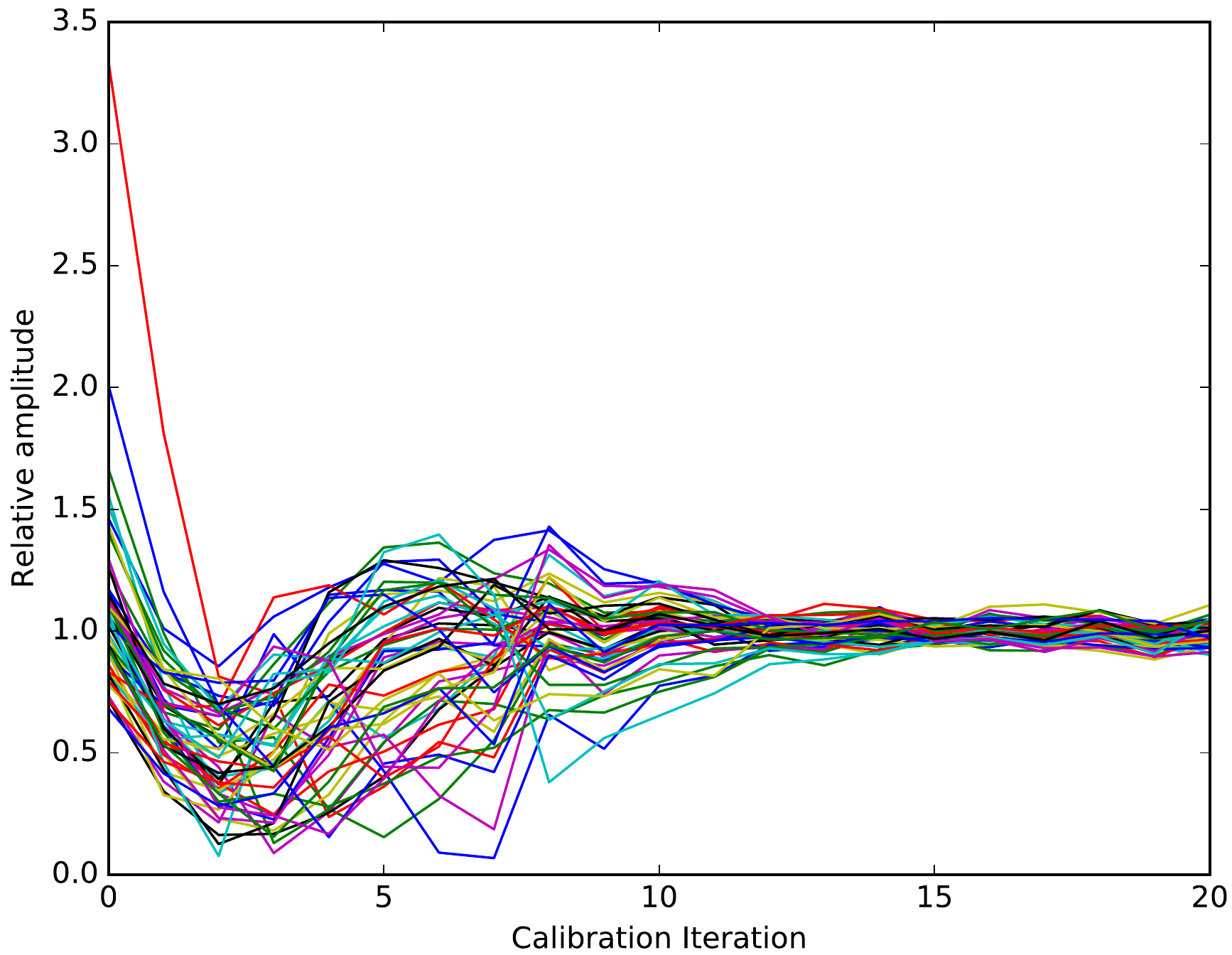}
\caption{Same as Fig.~\ref{fig:sim_phase}, but for gain amplitudes.
Each line represents an antenna in the 51 MWA antenna sample. 
The relative amplitude plotted is the magnitude of the estimated gains divided by the true gains.
After about 12 iterations we see the 
calibration loop has settled around the correct values, with only noise remaining.}
\label{fig:sim_amp}
\end{center}
\end{figure}

The estimated gain amplitudes for the simulations are shown in Fig.~\ref{fig:sim_amp}. The 
quantity plotted is the magnitude of the estimated gains over the true gains, $\left|g^{(n)}_a/
g_a\right|$, which places all antennas on the same scale. We can see the amplitudes 
converge toward their true values around the same time as the phases (iteration $\sim$12). At 
the beginning of calibration we can see the importance of the damping factor. At $n=0$, a couple of 
gains are shown to have abnormally high amplitude estimates, notably one about 3.3 times its 
true value (red line). These unbalanced high estimates caused the entire set of gains to be 
under estimated at $n=1$, even with a damping factor of 0.35. By $n=5$ the unbalanced 
amplitudes have been damped out and the calibration continues. Without the damping factor, 
the oscillation seen in the first couple iterations would have been significantly larger and taken 
much longer to fade out. 

After the gains have converged, we see both the phases and amplitudes continue to 
fluctuate coherently. This is due to the stochastic fluctuations of the sources 
themselves, which are simulated as zero mean random processes with variance proportional
to their flux density.
For this simulation we restricted each calibration 
iteration to only 10~ms integration time. In principle a calibration implementation could use 
short integration times to allow the gains to converge, then increase the integration time to 
reduce this noise.

\begin{figure*}
\begin{center}
\includegraphics[width=.3\linewidth]{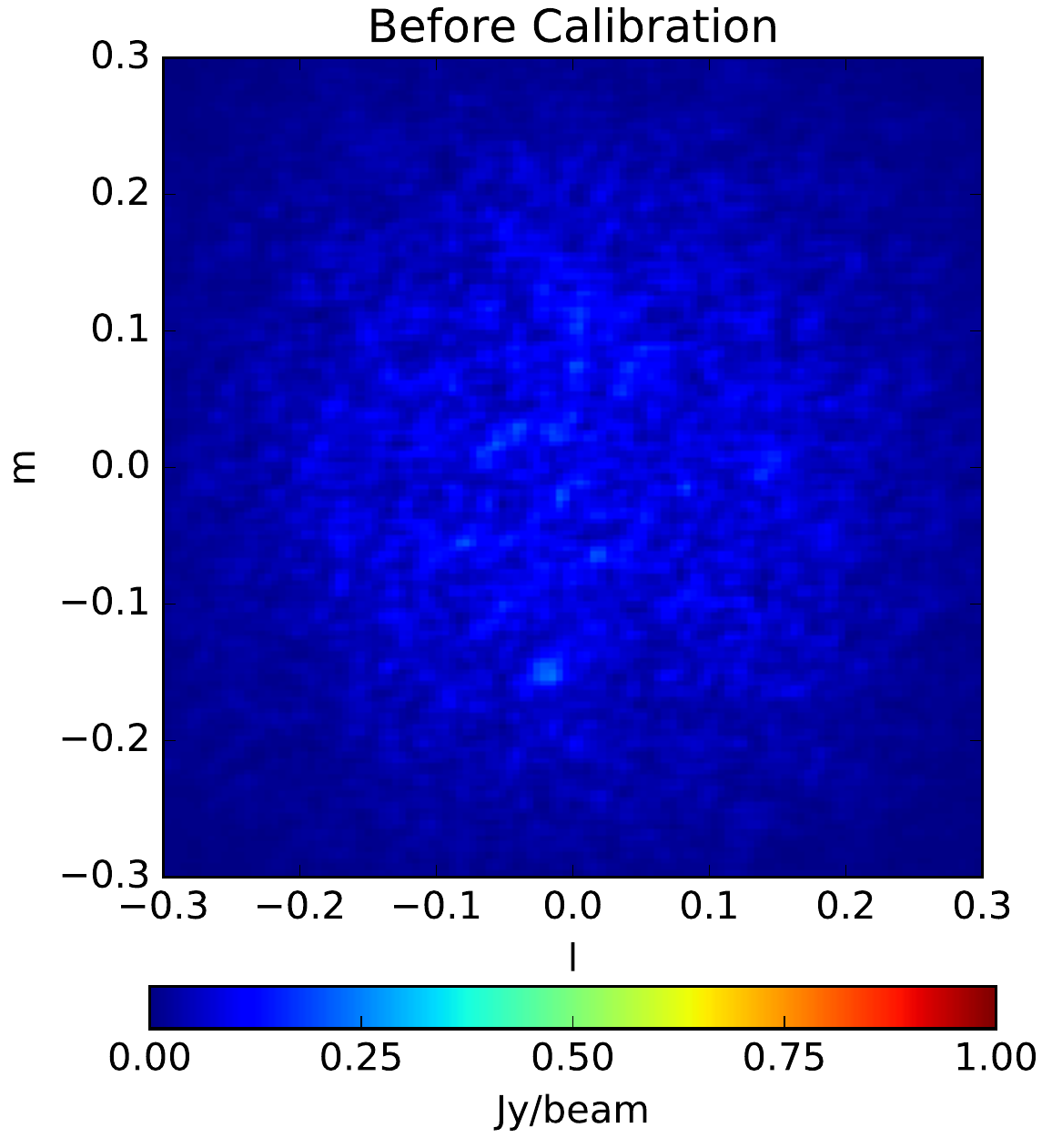}
\includegraphics[width=.3\linewidth]{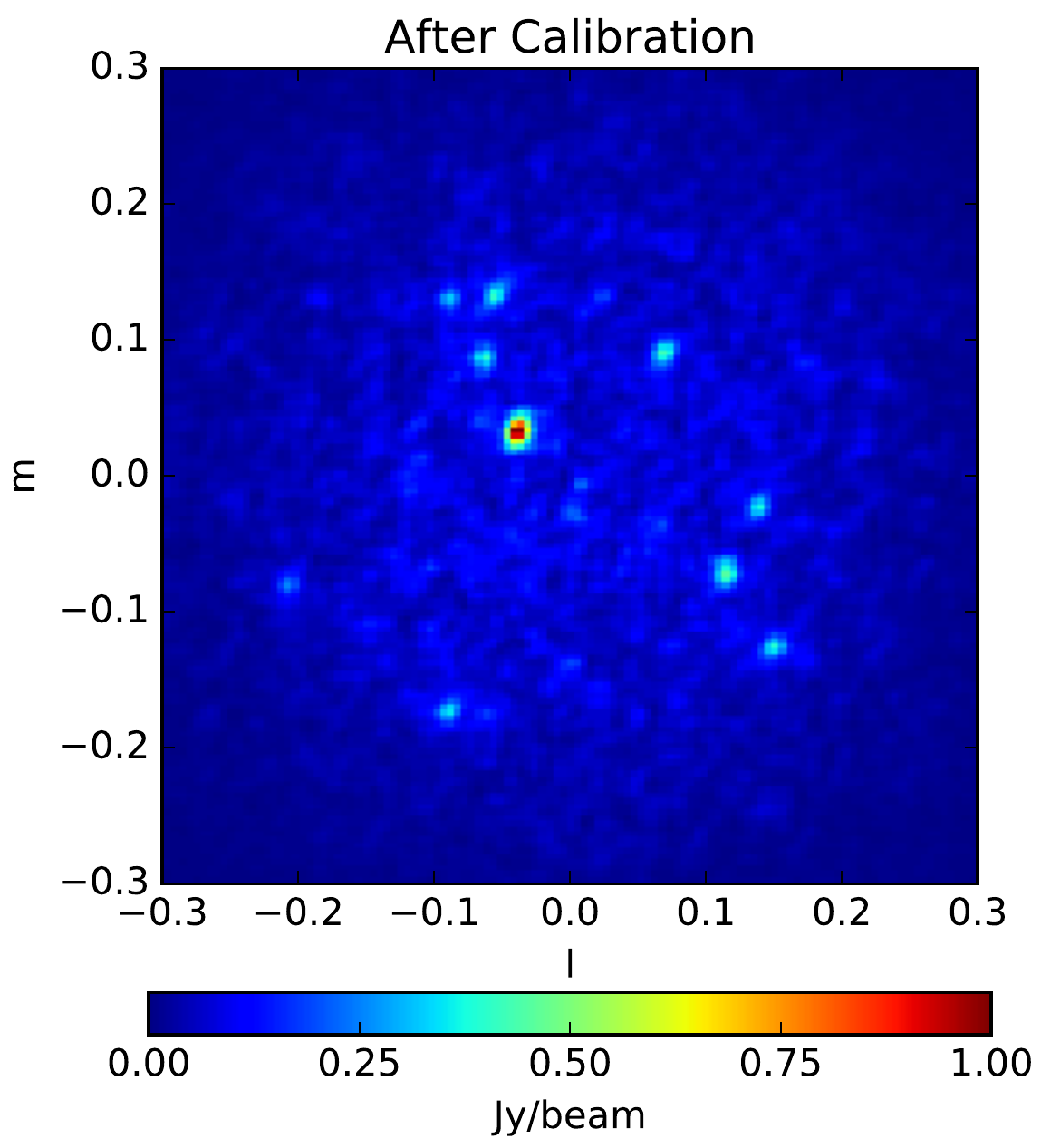}
\includegraphics[width=.3\linewidth]{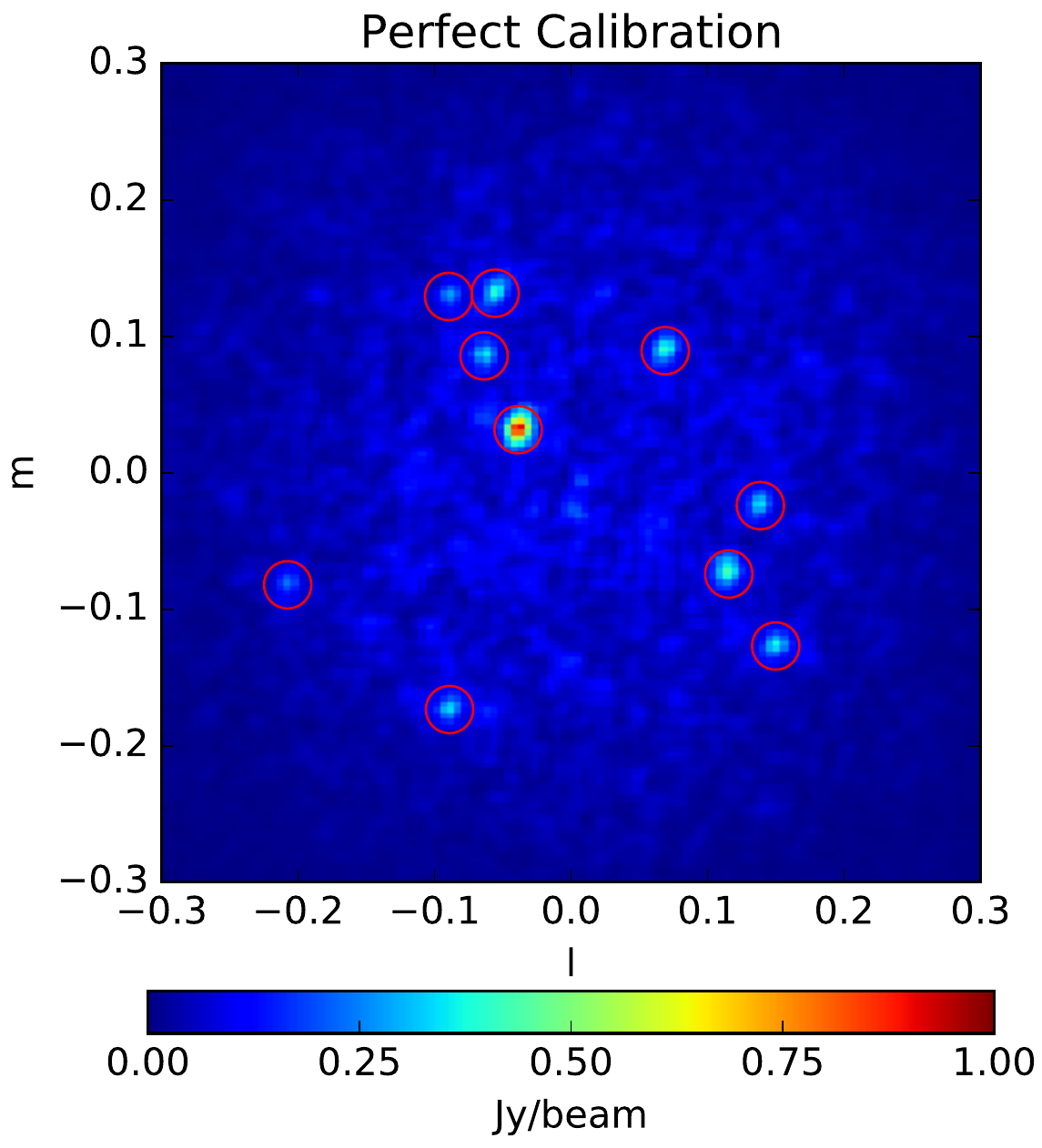}
\caption{Images formed during simulated calibration. \emph{Left:} An image generated from a 
single frequency channel and 400 time sample (10 ms) integration after the gain estimates are randomized. As 
expected with random phases, the image is completely noisy, with the shape of the primary 
beam evident. \emph{Middle:} An image formed after calibration, again with a single frequency 
channel and 400 time sample integration. Now the simulated point sources are visible in the image. 
\emph{Right:} An image generated from the same simulated data as the middle image, but 
with perfect calibration correction. Many of the features seen in the middle image are also
seen here, confirming that they are point spread function artifacts. The ten simulated (and modeled) point sources are highlighted with red circles.
\label{fig:sim_images}
}
\end{center}
\end{figure*}

Images created at the beginning of calibration and at the end are shown in 
Fig.~\ref{fig:sim_images}. Each image is obtained over a 400 time sample (10 ms) integration, corresponding to 
all snapshot images created with a given set of gain estimates. The left panel shows the image 
produced with our initialized unity gains. Because the phases are completely random, the 
image is essentially noise with the primary beam evident. After 20 calibration iterations, the image is far 
more clear, shown in the middle panel. 
The ten sources are visible, along with rumble throughout the image. By comparing with
a perfectly calibrated image (right panel), we see that the rumble and the shapes of the
point sources are matched between the two images. We therefore conclude that the 
artifacts visible are dominated by the point spread function, rather than calibration errors.

\section{Noise trends}\label{sec:noise}
Next we study the effect of noise in our system, and the consequences 
of an incomplete sky model. We run a suite of simulations varying the receiver noise and 
integration times, while forming antenna-pixel correlations, $\Cna$, for EPICal gain solutions, 
and simultaneously forming visibilities from the same streams of measured electric fields to find visibility-based 
gain solutions for comparison. 

We use a simulated sky consisting of a 5~Jy calibrator source, and 49 other random sources 
with apparent flux densities 0.2 to 0.5 Jy (total sky power $\approx$ 23 Jy). We generated 
these sources randomly, but kept them fixed for each run. 
We simulate the same antenna layout as in Section~\ref{sec:sim}. 
However, we increase the number of frequency channels simulated to 64, while keeping the flux of all sources constant across frequency.
Because our calibration loop treats 
frequency channels independently, each channel can be treated as a separate trial of the 
simulation, and is used to better estimate the statistics.

To simulate receiver noise, we add a gaussian distributed complex random number to each 
antenna electric field measurement at each time sample according to 
equation~\ref{eq:apply_gain}. The level of the receiver noise, $\sigma_r = \left<\left|
n_a(f,t)\right|^2\right>$, is varied in different simulation runs. We include two limiting 
cases, one where the receiver noise power is subdominant to the sky power ($\sigma_r = 
10.0$~Jy), and the other where the receiver noise dominates the total power ($\sigma_r=100.0$~Jy).

For each simulation run we find gain solutions using four methods:
\begin{enumerate}[i.]
\item EPICal, using a full sky model
\item EPICal, using a single point source model
\item Visibility-based, using a full sky model
\item Visibility-based, using a single point source model.
\end{enumerate}
In the cases where we use the full sky model, the model visibilities, $V^M_{ab}$, used in the 
calibration loop are created using all point sources in the simulated sky. For the single point 
source model, we only include the bright calibrator source in the model visibilities.
While we vary the model used for calibration, the simulated ``true sky'' remains fixed with all 50 
point sources.

For EPICal, we initialize our gain estimates with the true values 
(${g^{(0)}_a=g_a=1}$), and allow the estimate to be corrupted by the noise through ten 
iterations of the calibration loop. 
This allows us to bypass the initial convergence time and quickly study the effect of the
noise over many simuations.
We again adopt a damping factor $\damp=0.35$. Because EPICal 
updates the gain estimate at each calibration loop, \tcal, but retains a memory of previous 
iterations through the damping factor, the total integration time is not straightforward. We define 
an effective integration time by considering the relative weights of each previous $\Cna$ 
contributing to the current $g^{(n)}_a$ estimate, each with integration time \tcal. In the limit $n
\rightarrow \infty$, the geometric series converges to an effective integration time of
\begin{equation}\label{eq:teff}
\teff = \tcal \times \frac{1+\damp}{1-\damp}.
\end{equation}
With the damping factor we adopted here, $\teff \approx 2.1 \times \tcal$. 

We simultaneously form simulated visibilities, $V_{ab}$, by correlating all pairs of antenna electric field 
measurements (including the receiver noise). The electric field measurements are correlated for a duration equal to 
the effective integration time of the EPICal loop for comparison. We then find the 
visibility-based gain estimates by minimizing
\begin{equation}\label{eq:vis_cal}
\chi^2 = \sum_a\sum_{b\ne a} \left|V_{ab}-g_a g_b^* V^M_{ab}\right|^2,
\end{equation}
using both versions of the model visibilities described above.

 For each version of calibration, we observe the error in the gain estimates by averaging over both 
 antennas and frequency channels.
\begin{equation}\label{eq:gain_error}
\sigma_g = \left[\frac{1}{N_f \Nant} \sum_f \sum_a \frac{\left|g^{(n)}_a(f)-g_a(f)\right|^2}{\left|g_a(f)\right|^2}\right]^{1/2}
\end{equation}

\begin{figure}
\begin{center}
\includegraphics[width=\columnwidth]{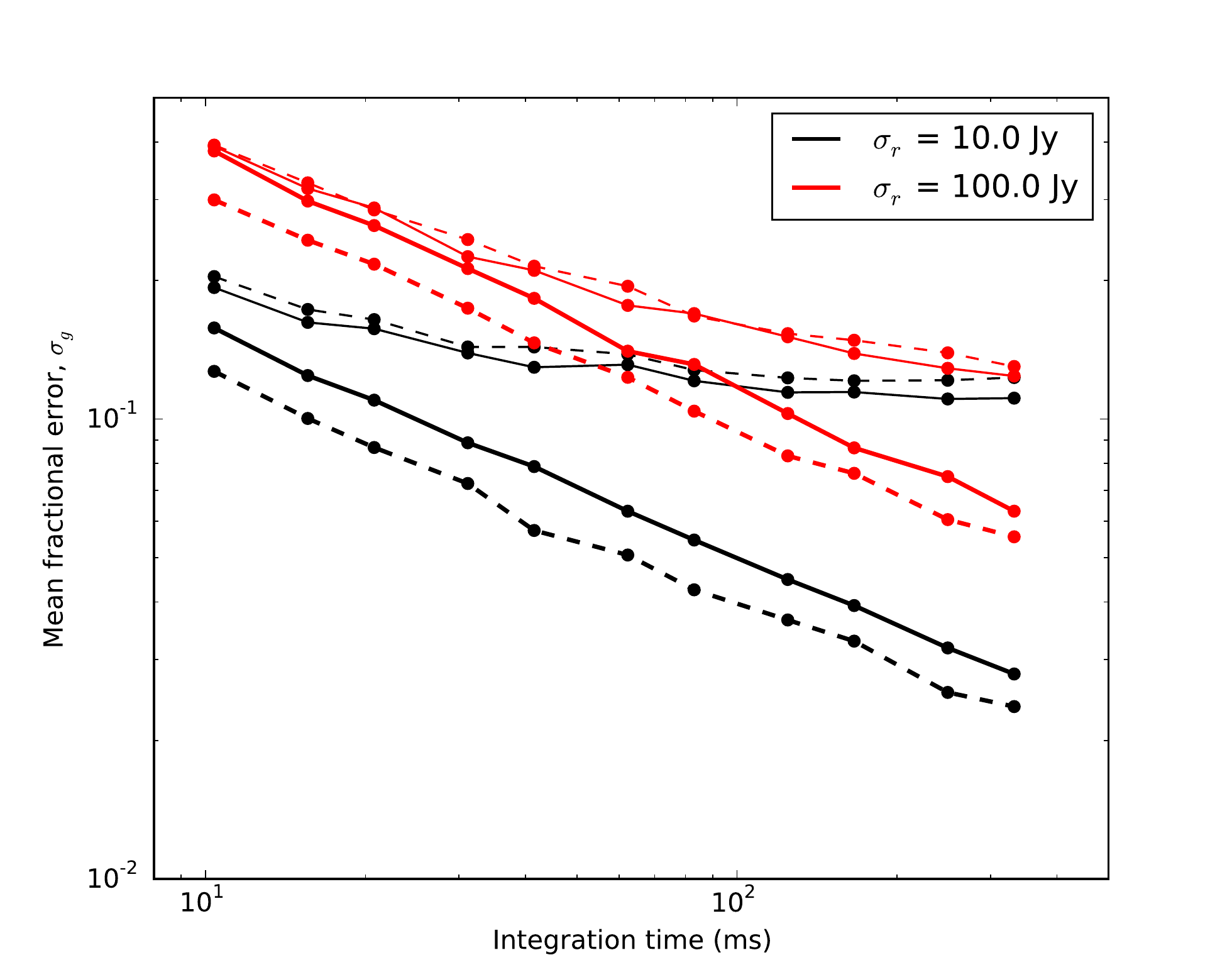}
\caption{Gain estimate errors as a function of integration time, receiver noise (line color), and 
calibration method. The solid lines represent the error on the EPICal derived gain estimates, 
while the dashed lines represent the error on the visibility based estimates. The thicker lines 
were derived using a full sky model in the calibration loops, and the thinner lines used only a 
single point source model. In the full sky model case we see all methods trend down with 
longer integration time as expected. For a given integration time, EPICal derived errors are on 
average 23\% higher than those of visibility based calibration. In the case of a single source 
model, the errors bottom out due to the flux in the measurements that is not modeled. In this 
regime the EPICal method performs slightly better than the visibility-based method due to 
effectively beamforming to the sky pixel where the incomplete model is most accurate. On 
average EPICal errors are 5\% lower in this regime.
The results from the high and low receiver noise regimes are similar, with both approaching the same floor in the single source model case.
}
\label{fig:errors}
\end{center}
\end{figure}

The results of our simulations for all four calibration methods are shown in Fig.~\ref{fig:errors}. 
In the case of the full sky model, we see the errors trend 
downward with longer integration times, as expected. The EPICal errors are slightly higher than 
the visibility-based errors, on average about 23\% difference. This is not surprising as EPICal 
only uses the information in a single pixel, while the visibilities use all information from the full 
sky model. However, we see that with this perfect model the errors do behave like noise
in the sense that they integrate down with longer integrations with the expected $t^{-1/2}$ 
trend. 
With longer integration time (or larger damping factor), EPICal can achieve the same 
level of gain error as visibility based solutions.

The exact ratio of noise in gains from EPICal and the visibility-based estimator
depends on the fraction of 
the total sky power contained in our calibrator source. In the example here, our calibrator 
accounted for 22\% of the total sky power. We repeated the experiment in the regime where 
the calibrator dominated the sky and found that the difference in EPICal and visibility-based 
error goes to zero, as expected. For a typical HERA observation, the sky temperature is 
expected to by about 180~K \citep{jac15}, and a bright calibrator source could be about 10\% 
of this power. In this regime we found that EPICal gain noise was about 60\% higher than 
visibility-based gain noise for the same effective integration time.

In a more realistic situation, the observer will not have a perfectly complete model of the sky.
A common method is to instead model the brightest sources in the field,
or even a single dominant source.
When using the single point source model, all calibration methods in Fig.~\ref{fig:errors} trend 
downward until they reach an error floor due to the confusion sources that were not modeled. 
The EPICal and visibility-based solutions reach a similar floor, but EPICal achieves a 
marginally lower level (on average 5\% lower). This can be attributed to the same reason 
EPICal underperformed with the full sky model: EPICal only used a single pixel in the sky to 
form its solutions. When using the full sky model, it was under utilizing the information in the 
rest of the sky. But when the sky model only contains the bright point source, the single pixel 
used in EPICal is the location where this model is most accurate, effectively down weighting 
the pixels with incomplete sky model. 
In contrast, the visibility-based calibration method 
does not have any preference to a certain direction, and
weights the sky equally (beyond the inherent response
of the instrument).
This results in a more harsh penalty for missing sources in the model.
In the regime typical of HERA discussed above, EPICal's 
error floor was 10\% lower than that of visibility-based solutions.

Any realistic sky model will lie somewhere between the two extremes explored here. In the 
case where the sky is modeled by a single source, as is often the case for an initial calibration, 
EPICal actually achieves smaller gain errors compared to the traditional visibility-based 
calibration. As the sky model improves, both visibility-based and EPICal gains improve, though 
the former quicker than the latter. In the limit of a perfect model, both calibration methods 
produce noise-like errors in their gains which scale down with more integration time.

Comparing the two regimes of receiver noise levels (red versus black lines in 
Fig.~\ref{fig:errors}), we see that the gains errors behave as one might expect given our
above interpretations. In the case of a perfect sky model, the higher receiver noise results
in larger gain errors for a given integration time, but follows the same $t^{-1/2}$ noise 
trend. The imperfect model lines begin exhibit the same offset at low integration time
because the errors are dominated by the thermal noise. But as integration time increases,
they approach the same floor due to the same level of power from unmodeled sources.

\section{Application to LWA data}\label{sec:data}
We next demonstrate our calibration algorithm using an observation from the LWA station in 
New Mexico \citep{ell13}. The data is from the LWA narrow-band transient buffer (TBN), with time domain
voltage data from the 255 core antennas within a 110~m x 100~m ellipse. 
The central frequency is 74.03 
MHz, with a 
total observing bandwidth of 100~kHz, analyzed by a filter bank into 512 channels of bandwidth 195.3125~Hz each and with a post-channelization sampling period of 5.12~ms.
For this demonstration we limit ourselves to a single polarization.

For the gridding step, we adopt a grid with extent 120~m to encompass the footprint 
of the array, and a grid spacing of half wavelength ($\approx 2$~m). The result is a 
$64\times64$ pixel grid, making the MOFF computational complexity of 
$\Ng \log_2 \Ng$ on the same order as the $\Nant^2$ computations for a cross correlator. 
Our gridding kernel is a square tophat function with sides of 3~m, corresponding to the size of the LWA
ground screens.

After correcting for geometric cable delays, the instrument is sufficiently calibrated to produce recognizable images of the sky, as was 
seen in the demonstration of the EPIC imager in \citealt{thy17} (their section 4.4). However, we will aim to 
further improve on this calibration using our algorithm.

We proceed by forming model visibilities. We model only two bright objects as point sources: 
Cyg A with flux 20,539 Jy, and Cas A with flux 19,328 Jy \citep{lan12}. 
Because the raw data are attenuated by the primary beam of the instrument, we also account for 
this in our model using beam values consistent with \cite{hic12}.

We made several choices while studying the behavior of the LWA data to improve our 
calibration. Through our previous imaging work, we noted that the flux scale of uncalibrated 
images was consistent with average gain amplitudes of roughly 0.25. To allow the calibration to 
converge quickly we initialized our gain estimates at this level. 
We also found a boost in signal to noise ratio is achieved easily by assuming the gains are
constant across a range of frequencies, and averaging solutions across channels.
Here we average solutions across the central 300 channels, or about 58.6 kHz, 
and discard the remaining channels which are near the edge of the observational
band and have significantly different gain amplitude.
With a fractional bandwidth $B/f_0 = 7.9 \times 10^{-4} \ll 1$, we assume a smooth bandpass 
across the band. 
A damping factor of $\gamma = 0.7$ was adopted.
We found a larger damping factor, relative to what was used in simulations, was
beneficial with the LWA data. Several factors likely contribute to the difference in damping
factors, including increased thermal noise in the LWA data compared to the simulations,
as well as different antenna designs, array configurations, and sky signals.
A larger damping factor yields a larger effective integration time and thus lower noise on
each calibration solution (Eq.~\ref{eq:teff}). We save for future work an investigation of robust 
determination of optimal damping factors.

\begin{figure}
\begin{center}
\includegraphics[width=\columnwidth]{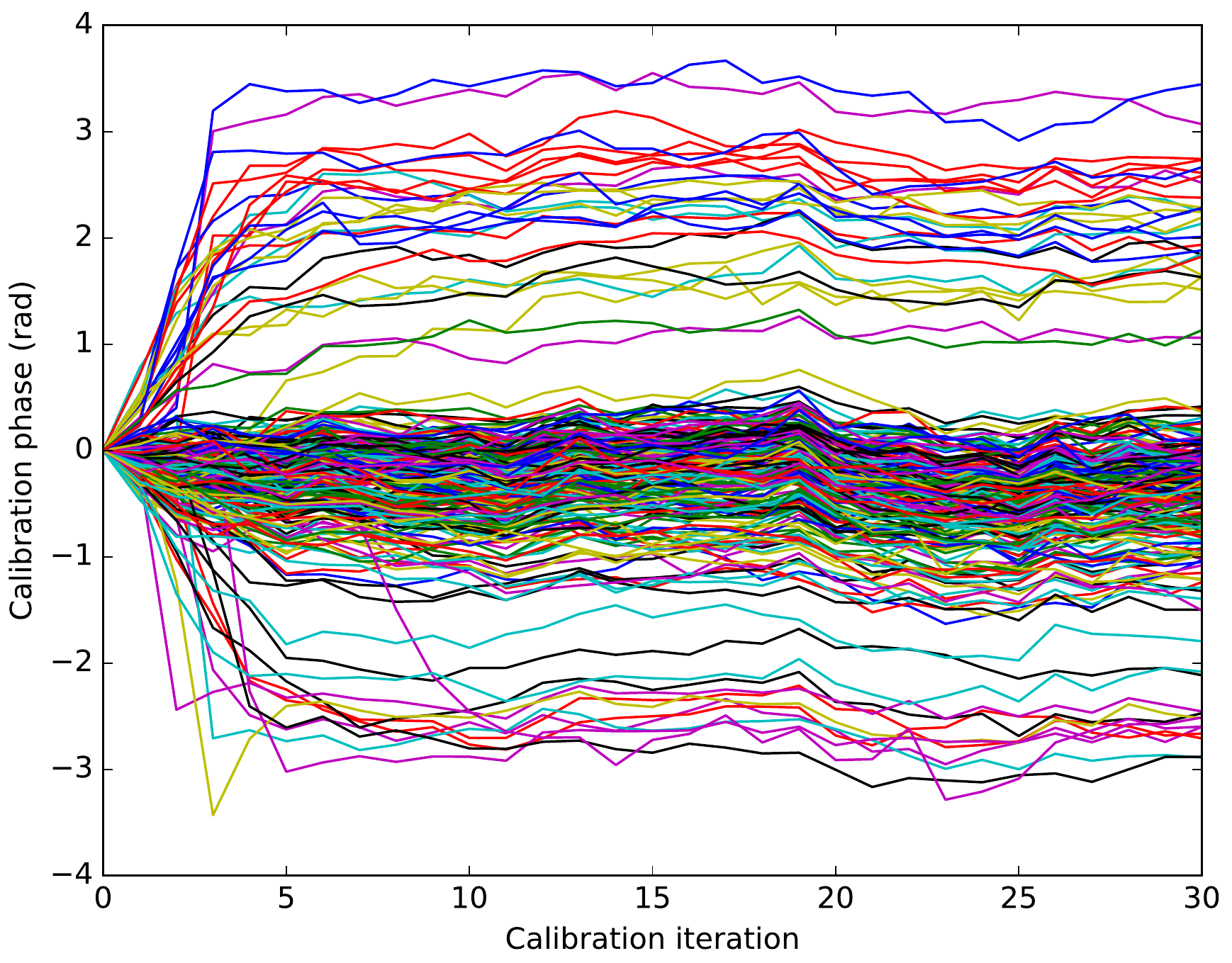}
\caption{Gain phase solutions as a function of calibration iteration for an LWA TBN observation. 
The gain estimates are initialized with zero phase, but quickly span a $2\pi$ range, and settle 
into relatively flat, albeit noisy, solutions. The majority of phases congregate near zero, which is 
not surprising given the fairly good quality image produced from uncalibrated data.
For plotting clarity, we unwrapped phases resulting in phases that appear to exceed 
$\pm \pi$.
}
\label{fig:data_phase}
\end{center}
\end{figure}

\begin{figure}
\begin{center}
\includegraphics[width=\columnwidth]{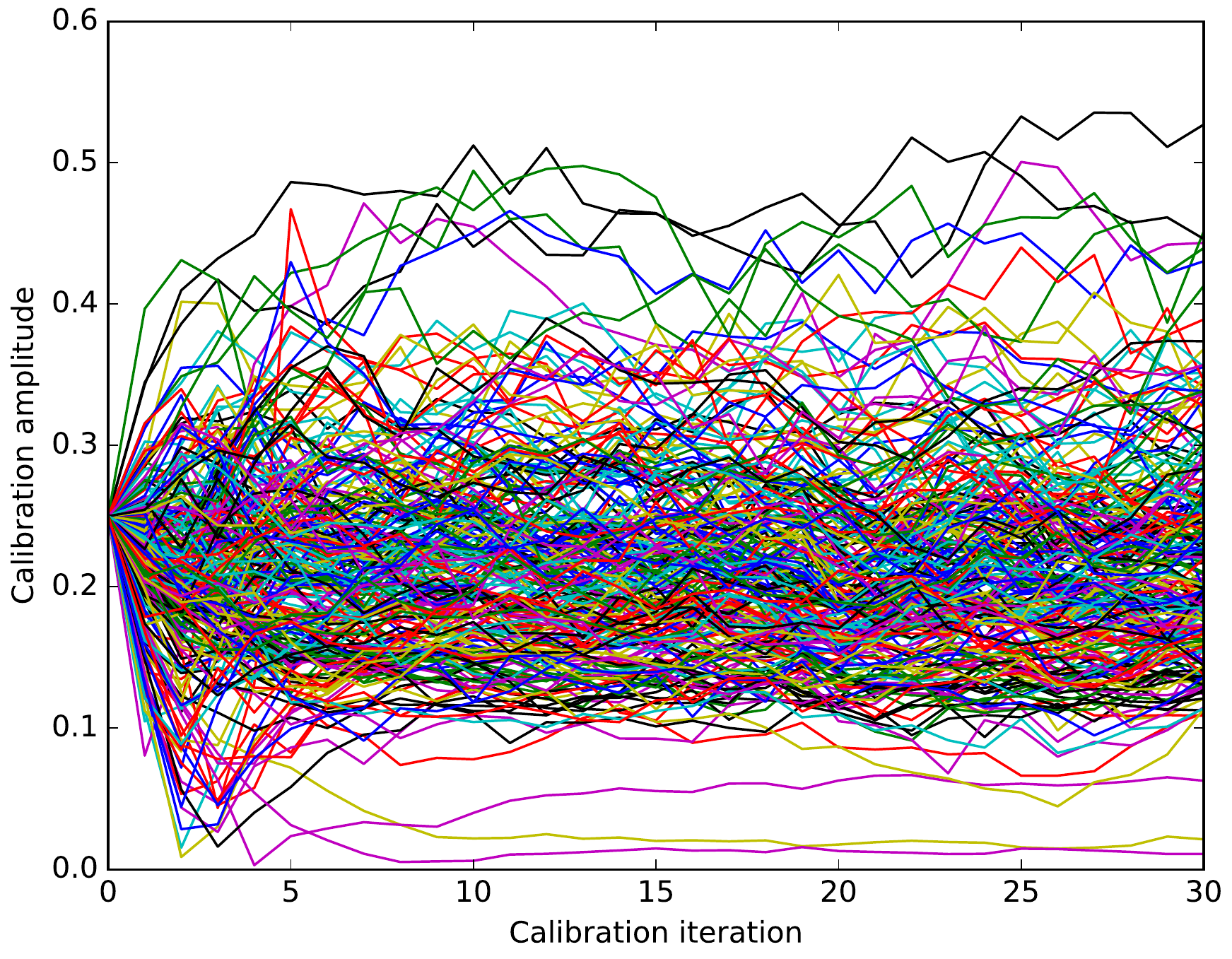}
\caption{Same as Fig.~\ref{fig:data_phase}, but for gain amplitudes.
The 
gains estimates were initialized with amplitude 0.25 after inspection of the electric field values 
compared to our model sky. The solutions are noisy, but flat. The range in amplitudes is due to 
the non-uniformity of cables between the LWA antennas and receivers.
}
\label{fig:data_amp}
\end{center}
\end{figure}

\begin{figure*}
\begin{center}
\includegraphics[width=0.3\linewidth]{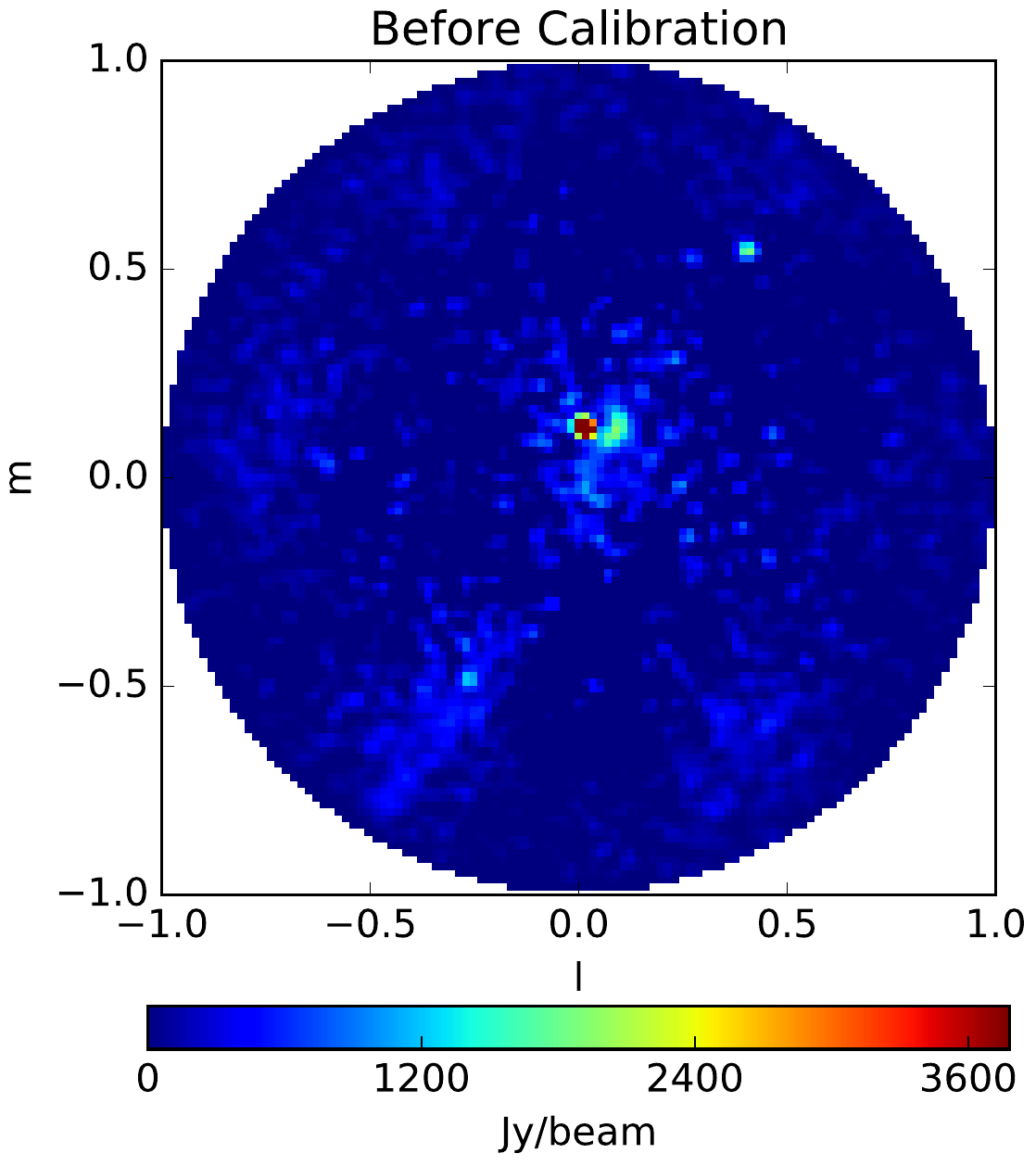}
\includegraphics[width=0.3\linewidth]{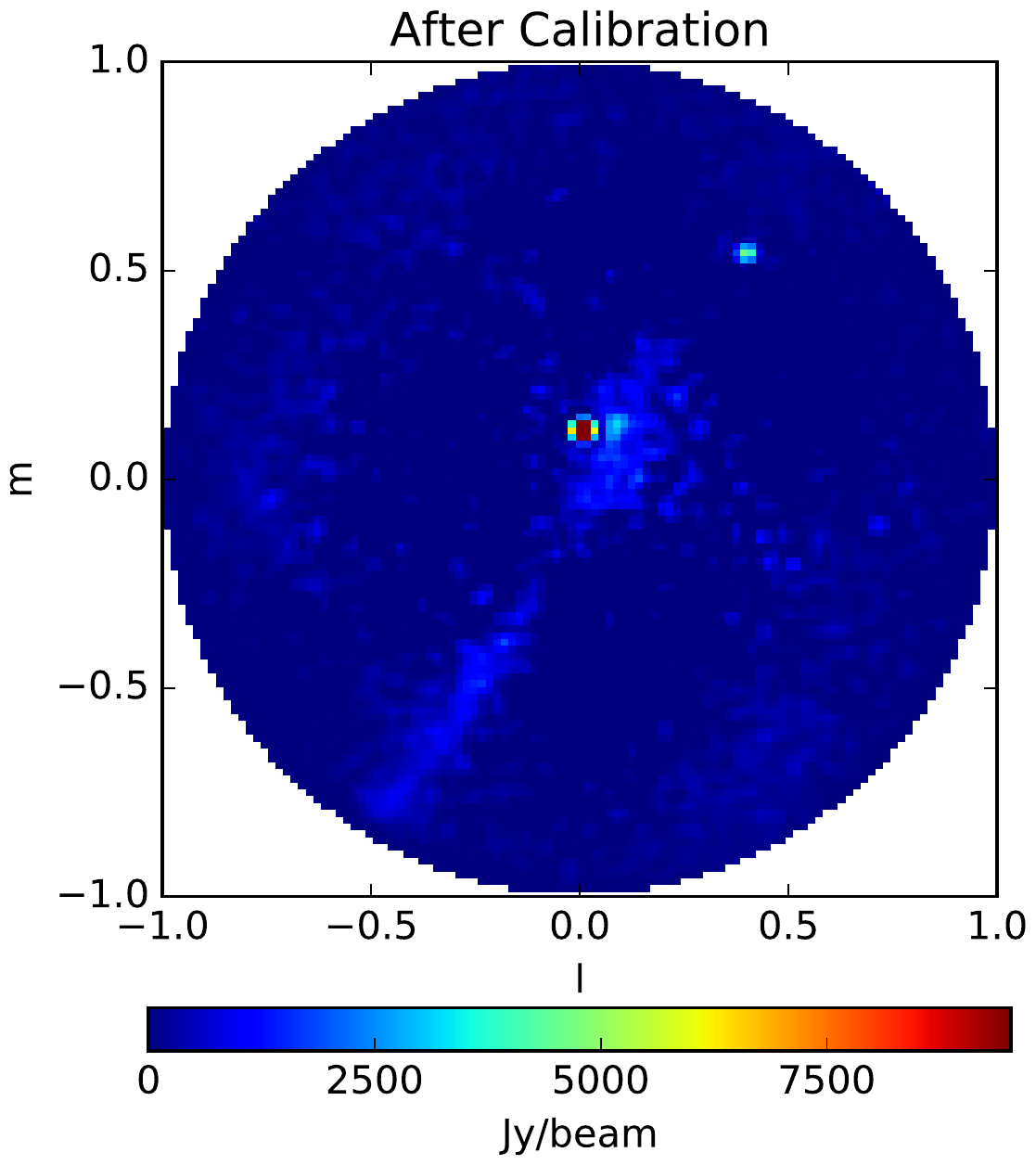}
\includegraphics[width=0.3\linewidth]{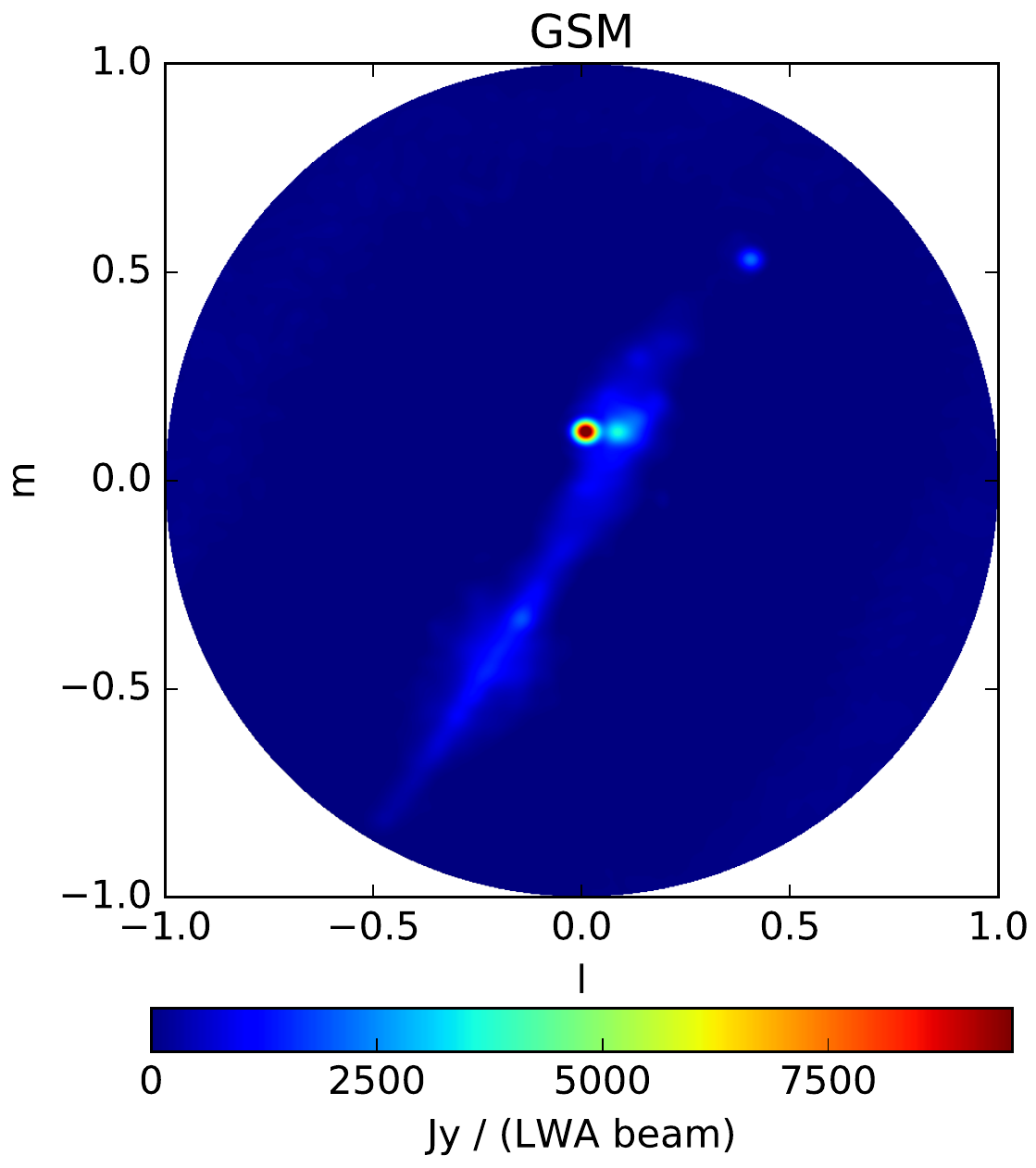}
\caption{Images produced before (left) and after (middle) calibrating LWA data. These images 
were produced with 58.6 kHz bandwidth and 51.2 ms integration. 
The color scales are saturated at half the peak of the respective images.
The calibrated image shows 
significant reduction in rumble throughout, while retaining the prominent Cyg~A and 
Cas~A sources. The galactic plane is also substantially more evident after calibration. For 
reference, the GSM is shown in the 
right panel convolved with the LWA point spread function and weighted by two factors
of the primary beam to place it in the same holographic frame as the data images.
(One beam factor is from the instrument measurement, and the other is from the
gridding step of the analysis.)
While much of the GSM is visible in the calibrated image, we 
note that the model used to calibrate was only the two bright sources, Cyg~A and Cas~A.
}
\label{fig:data_images}
\end{center}
\end{figure*}

Finally, we found that seven antennas\footnote{LWA antenna IDs 48, 85, 124, 148, 203, 217, 
and 244 were omitted.} produced unstable gain solutions, and in fact corrupted the entire array. 
We therefore omitted these antennas from our analysis, resulting in a total of 248 antennas to 
calibrate. In each calibration loop, we form $C_{i,\hat{\boldsymbol{s}}_0}, i=1,2,\ldots N_a$ and 
update our gain estimates over 10 samples (51.2 ms). We iterate the loop 30 times for a 
total of 1.536 seconds of data processed. The results of this calibration experiment are shown in Figs.~
\ref{fig:data_phase}~--~\ref{fig:data_images}.

Figure~\ref{fig:data_phase} shows the phase of our gain estimates over 30 calibration 
iterations, again with each colored line representing a different antenna. Given the quality of 
uncalibrated image demonstrated in \cite{thy17}, we had anticipated that the phase
solutions would not exhibit much variation between antennas. Hence the relatively large variation was unexpected.
However, the phases are relatively flat after about 15 iterations 
(modulo noise), and exhibit a central ``trunk" where the majority of phases are congregated. 
This behavior is suggestive that while the uncalibrated data were able to produce a viable 
image, the minor changes from our solutions will focus the image and improve the quality. The 
actual location of the ``trunk" (slightly negative) is simply determined by the reference antenna 
chosen to have identically zero phase, but happens to be slightly more positive than the bulk of 
antennas. 

The gain amplitudes as a function of calibration iteration are shown in 
Fig.~\ref{fig:data_amp}.
To ensure the gain solutions and noise levels are reasonable, we compared to
solutions found with a visibility-based approach. We first form visibilities by directly correlating
antenna electric field measurements within each calibration iteration time interval. We then
solve for visibility-based gains for each antenna and calibration iteration by minimizing
$\chi^2$ in equation~\ref{eq:vis_cal}. Because the field has significant diffuse structure 
from the Galactic plane, we omit baselines shorter than 10 wavelengths. We then use an
initial gain estimate of 0.25 (as we did above for EPICal), and apply the damping factor at
each successive calibration iteration to achieve a weighted average analogous to the EPICal 
solutions.
The resulting gain phases spanned the full $2\pi$ range, exhibiting the same ``trunk''
we see in Fig.~\ref{fig:data_phase}. The gain amplitudes spanned a wide range similar to
our EPICal results (from 0.004 to 0.68).

We next compare the noise levels between our EPICal solutions and visibility-based 
solutions. We estimate the fractional gain noise for each method by calculating the standard 
deviation of the gain estimates after iteration 15 for each antenna separately, and dividing by 
the respective mean gain amplitudes. For the EPICal solutions we found a fractional gain
noise of 0.15, and for the visibility-based solutions we found 0.12. We conclude that in this 
demonstration, the fluctuations seen in Figures~\ref{fig:data_phase} and \ref{fig:data_amp}
are about 25\% larger than we see for a visibility-based method, which is consistent with
the noise trends we saw in Section~\ref{sec:noise} when noise dominated the error.

Figure~\ref{fig:data_images} shows the improvement in the images due to our calibration. The 
left panel shows the uncalibrated image integrated over 51.2 ms, 58.6 kHz. Cyg~A is prominent 
near the center of the image, and Cas~A is also clearly visible in the upper right. The middle 
panel shows the image produced after calibration with identical integration time and bandwidth. 
We show each of these images with a color scale which saturates at half the
maximum value in the respective image data. We find that this choice allows us to
qualitatively compare the quality of the images, independent of the absolute scale which
changed during the calibration.
The sidelobes throughout the calibrated image are significantly suppressed, and the galactic plane is 
much more evident, despite only modeling Cyg~A and Cas~A. We also note that the feature 
just to the right of Cyg~A is dimmer in the calibrated image, better matching the expected flux 
from the GSM. 
For reference we show the GSM in the right panel of 
Fig.~\ref{fig:data_images}, convolved by the LWA point spread function and weighted
by  two factors of the primary beam to match the holographic frame of the data images.

To compare the image qualities quantitatively we compute the dynamic range, defined as the 
peak of the image over the noise level. We estimate the noise level as the median of the 
absolute deviation of the image. With this metric we find the uncalibrated 
and calibrated images to have dynamic ranges of 80.4 and 133.1, respectively, a 65\% 
improvement.

\section{Discussion}\label{sec:discussion}
Through simulations and application to real data, we have shown that the EPICal algorithm is a 
viable solution for calibrating direct imaging arrays in real time. The computation necessary 
only scales with the number of antenna elements, making it a sub-dominant cost factor when 
designing the correlator. This strategy will enable fast read-out for arrays with many thousands 
of antennas, which will be necessary for future radio transient and cosmology experiments.

EPICal can be further improved through several extensions. Here we name a few potential 
considerations for further study. These topics can be thought of as extensions to the white 
``estimate gains" box in Fig.~\ref{fig:schematic}, which can be performed at a much 
slower rate than the actual antenna-pixel correlations.

\textbf{Multiple pixel correlations.} As was seen in section~\ref{sec:noise}, the single pixel 
correlation demonstrated in this paper can underperform in the presence of a complex sky, 
when compared to visibility-based gains with perfect knowledge of the sky. These errors can be 
mitigated by using correlations of multiple image pixels to incorporate a higher fraction of the 
total sky power into the calibration loop. Of course this increases the computational cost to a 
scaling of $\mathcal{O}(\Nant N_{\mathrm{pix}})$, which will typically be much lower than the 
$\mathcal{O}(\Ng \log_2 \Ng)$ of the correlator itself. These additional correlations would also 
enable direction dependent gain solutions. Each correlation could be used to independently 
solve for gains in each pixel direction, then fit to a beam model on the sky. This updated beam 
pattern would then feed into the gridding step of the correlator, allowing the imager to convolve 
the signals with the effective beam pattern in the ground plane. 

\textbf{Fitting gain models.} With some knowledge of the instrumental bandpass, the noise of 
the gain solutions can be greatly reduced by fitting a model to the per-frequency solutions 
derived here. We used this method at a rudimentary level in our demonstration to LWA data by 
assuming the gains were constant over a narrow bandwidth. One could easily improve on this 
by extending the bandwidth and fitting for a low order polynomial in phase and amplitude. 
Additionally, with knowledge of 
any filters applied to the data stream in the receiver chain,
we could include channels closer to the edge of the band.

\textbf{Improving sky model.} In this work we used a perfect sky model (for simulation), or a 
very simple sky model (for LWA data). In principle the images produced by the EPIC correlator 
can be used to improve the sky model used in calibration -- similar to a major loop in 
self-calibration. This can be especially useful for compact, widefield arrays which require a model of 
both compact and diffuse sources over a large patch of sky. 
As with any sky-based calibration scheme, EPICal requires that the sky model be attenuated
by the primary beam, which 
can be difficult to measure at the precision necessary \citep[e.g.][]{neb15,vir14,thy15b}. 
However, the direct imaging correlator provides exactly the product of the sky and beam necessary, and 
can be iterated over to improve the images and gain solutions, leaving final beam
correction to post-processing as in traditional processing.

\textbf{Dynamic parameters.} In our controlled experiments we fine-tuned a number of 
parameters based on our testing (e.g. damping factor, integration time, frequency averaging). A 
deployed system will require robust determination of these parameters to operate continuously. 
The specifics will be heavily dependent on the stability of the instrument, the frequency of 
observation, and the sky. For example, a stable instrument may be able to use short 
integrations to determine a rough estimate of the gains before switching to much longer 
integration (on order seconds to minutes) to highly increase signal to noise. At low frequencies 
or high imaging resolution, the dynamics of the ionosphere are important, and will likely drive 
the limit of time integration allowed.

The work here will serve as a foundation for further development. We have shown that the 
EPICal algorithm produces reliable calibration solutions, and have identified several aspects to 
increase the scope. With next generation instruments in the planning and development stages, 
EPICal is poised to enable new design spaces. The software is integrated into the EPIC 
package and freely available to use in simulations or post-processing of data. Work is 
underway to port the code to GPU systems for deployment. 

\section*{Acknowledgements}
This work has been supported by the National Science Foundation through award 
AST-1206552. We thank Danny Jacobs for his valuable inputs, and Greg Taylor for providing 
us with LWA data. Construction of the LWA has been supported by the Office of Naval 
Research under Contract N00014-07-C-0147. Support for operations and continuing 
development of the LWA1 is provided by the National Science Foundation under grant 
AST-1139974 of the University Radio Observatory program.



\interlinepenalty=10000
\bibliographystyle{mnras}
\bibliography{../epic} 

\bsp	
\label{lastpage}
\end{document}